\documentclass[intlimits,twoside,a4paper]{article}

\usepackage{amsmath,amssymb}
\usepackage{graphicx,psfrag}
\usepackage{wrapfig}
\usepackage[T2A]{fontenc}
\usepackage[cp1251]{inputenc}

\usepackage[eqsecnum]{cmpj}
\issue{2011}{14}{3}{33701}

\doinumber{10.5488/CMP.14.33701}

\date{Received May 11, 2011}

\authorcopyright{V. Blavatska, C. von Ferber, Yu. Holovatch, 2011}

\title[Shapes of macromolecules in good solvents:
field theoretical renormalization group approach]%
{Shapes of macromolecules in good solvents:\\
field theoretical renormalization group approach\thanks{A paper dedicated to
Prof. Yurij Kalyuzhnyi on the occasion of his 60th birthday.}}
\author[V. Blavatska, C. von Ferber, Yu. Holovatch]{V. Blavatska\refaddr{label1}, C. von Ferber\refaddr{label2,label3}, Yu. Holovatch\refaddr{label1}}
\addresses{
\addr{label1} Institute for Condensed Matter Physics of the
National Academy of Sciences of Ukraine, \\1~Svientsitskii Str.,
79011~Lviv, Ukraine
\addr{label2} Applied Mathematics Research Centre, Coventry University, CV1 5FB Coventry, UK
\addr{label3} Theoretische Polymerphysik, Universit\"at Freiburg,
79104~Freiburg, Germany
}

\begin{document}

\maketitle

\begin{abstract}
In this paper, we show how the method of field theoretical renormalization group
may be used to analyze universal shape properties of long polymer chains in porous
environment. So far  such analytical calculations were primarily focussed on   the
scaling exponents that govern conformational properties of polymer macromolecules. However,
there are other observables that along with the scaling exponents are universal (i.e.~independent of
the chemical structure of macromolecules and of the solvent) and may be analyzed within the
renormalization group approach.  Here, we address the question of shape which is acquired
by the long flexible polymer macromolecule when it is immersed in a solvent in the  presence of a
porous environment. This question is of relevance for understanding of the behavior of macromolecules
in colloidal solutions,  near microporous membranes, and in cellular environment.
To this end, we consider a previously suggested model of polymers in $d${-}dimensions
[V.~Blavats'ka, C.~von Ferber, Yu.~Holovatch, Phys. Rev.~E, 2001,  {\bf 64}, 041102] in an
environment with structural obstacles, characterized by a pair correlation function $h(r)$,
that decays with distance $r$ according to a power law:  $h(r) \sim r^{{-}a}$. We apply the
field{-}theoretical renormalization group approach and estimate the size ratio
$\langle R_{\rm e}^2 \rangle/\langle R_{\rm G}^2 \rangle$ and the asphericity ratio $\hat A_d$
up to the first order of a double $\varepsilon=4{-}d$, $\delta=4{-}a$ expansion.
\keywords polymer, quenched disorder, renormalization  group
\pacs 75.10.Hk, 11.10.Hi, 12.38.Cy
\end{abstract}

\section{Introduction}

Polymer theory belongs to and uses methods of different fields of science: physics, physical chemistry,
chemistry, and material science being the principal ones. Historically, the structure of polymers remained under
controversial discussion until, under the influence of the work by Hermann Staudinger~\cite{staudinger}, the idea of long chain{-}like
molecules became generally accepted.
In this paper we will concentrate on the {\em universal properties} of long polymer
chains immersed in a good solvent, i.e. the properties that do not depend on the chemical structure of macromolecules and of the solvent.
Usually a self{-}avoiding walk model is used to analyze such properties~\cite{deGennes79,Vanderzande}. At first glance
such a model is a rough  caricature of a polymer macromolecule since out of its numerous inherent
features it takes into account only its connectivity and the excluded volume modeled by a delta{-}like
self{-}avoidance condition. However, it is widely recognized by now that the universal  conformational properties of
polymer macromolecules are perfectly described by the model of self{-}avoiding walks.
It is instructive to note that the idea to describe polymers in terms of statistical mechanics
appeared already in early 30{-}ies due to Werner Kuhn~\cite{Kuhn34} and already then enabled
an understanding and a qualitative description of their properties. The present success in their
analytic description which has lead to accurate quantitative results is to a large extent due
to the application of field theoretic methods. In the pioneering  papers by Pierre Gilles
de Gennes and his school~\cite{deGennes79} an analogy was shown between the universal behaviour of spin systems
near the critical point and the behaviour of long polymer macromolecules in a good solvent. In turn,
this made it possible to apply the methods of field  theoretical renormalization group~\cite{rgbooks} to polymer
theory.

In spite of its success in explaining the universal properties of polymer macromolecules, the self-avoiding walk model does not encompass a variety of other polymer features. Different models and different
methods are used for these purposes. In the context of this Festschrift it is appropriate to mention
the approach based on the integral{-}equation techniques which is actively developed by Yurij
Kalyuzhnyi and his numerous colleagues~\cite{Kalyuzhnyi1}. In particular, this approach has enabled an
analytic description of chemically associating fluids and the representation of the most important
generic properties of certain classes of associating fluids~\cite{Kalyuzhnyi2}. It is our pleasure
and honor to write a paper dedicated to Yurij Kalyuzhnyi on the occasion on his 60th birthday and
doing so to wish him many more years of fruitful scientific activity and to acknowledge our numerous
common experiences in physics and not only therein.

In this paper we will analyze the shapes of  polymer macromolecules in a good solvent.
Flexible polymer macromolecules in dilute solutions form crumpled coils with a global shape, which greatly differs from spherical symmetry and is surprisingly anisotropic, as it has been found experimentally and confirmed
in many analytical and numerical investigations~\cite{Solc71,Rudnick86,Aronovitz86,Cannon91,Bishop88,Domb69,Jagodzinski92,Witten78,Kranbuehl77,Benhamou85,
Honeycutt88,Cardy89,Haber00,Dima04,Rawat09}.
Topological properties of macromolecules, such as the shape and size of a typical polymer chain configuration, are of interest in various respects. The shape of proteins affects their folding dynamics and motion
in a cell and is relevant in comprehending complex cellular phenomena, such as
catalytic activity~\cite{Plaxco98}. The hydrodynamics of polymer fluids is essentially affected by the size and shape of individual
macromolecules~\cite{Neurath41}; the polymer shape plays an important role in determining its molecular weight
in gel filtration chromatography~\cite {Erickson09}.
Below we will show how the shape can be quantified within universal characteristics  and how to calculate these characteristics analytically.

The set up of this paper is as follows. In the next section we present
some details of the first analytical attempt to study the shape of linear polymers in good solvent,
performed by Kuhn in 1934. Since then, the  study of topological properties of
polymer macromolecules was developed, based on a mathematical description, which is presented
in section~3 along with a short review of the known results for shape characteristics of flexible polymers. In section~4, we present details of the application of the field-theoretical renormalization group
approach to the study of universal polymer shape characteristics. Section 5 concerns the effect of structural disorder in the environment on the universal properties of polymer macromolecules. The model with long-range-correlated quenched defects is exploited, and the shape characteristics are estimated in a field-theoretical approach. We close by giving conclusions and an outlook.

\section{Shape of a flexible polymer: Kuhn's intuitive approach{\label{2}}}

\begin{figure}[t!]
\begin{center}
\includegraphics[width=8cm]{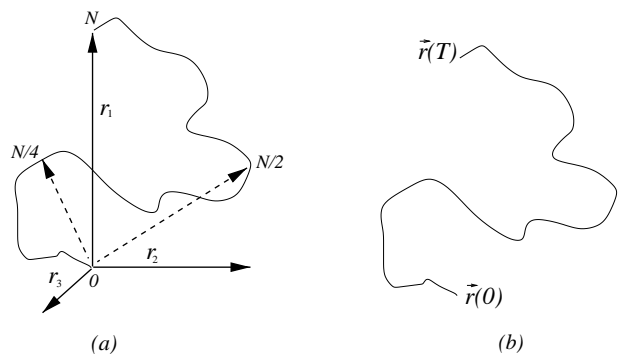}
\end{center}
\caption{\label{scheme}  a) Schematic presentation of a  polymer
chain with the end-to-end distance (position vector of $N$th
monomer) denoted as $\vec{r}_1$\,. The component of the position
vector of $N/2$th monomer in direction perpendicular to $\vec{r}_1$
is denoted by $\vec{r}_2$\,. Vector $\vec{r}_3$ denotes the component
of position vector of $N/4$th monomer in direction perpendicular to
both $\vec{r}_1$ and $\vec{r_2}$\,. b) In the Edwards continuous chain
model, the polymer is represented by a path $\vec{r}(t)$
parameterized by $0\leqslant t \leqslant T$ (see section~4). }
\end{figure}

The subject of primary interest in this paper will be the shape of a long flexible polymer
macromolecule. That is, assuming that a polymer coil  constitutes  of a large sequence of
monomers, does this coil resemble a globe (which would be the naive expectation taken that
each monomer is attached at random) or does its shape  possess anisotropy. And, if it is anisotropic,
what are the observables to describe it quantitatively? In this section we start our analysis with
a very simple model that allows us to make some quantitative conclusions. The  aim of
the calculations given below is to explain how the anisotropy of the polymer shape arises already
within a random walk model. That is to show that the anisotropy is essentially not  an excluded
volume effect (although its strength is effected by the excluded volume interaction as we will show
in this paper) but rather it is  an intrinsic property arising from random walk statistics.

The analysis given below is inspired by Kuhn's seminal paper~\cite{Kuhn34}. However, Kuhn's explanation is based on
combinatorial analysis and application of Stirling formula, while here we
suggest a derivation based on the application of Bayesian probability~\cite{Bayes}. Let us consider the so-called Gaussian freely jointed
chain model consisting of $N$ connected bonds capable of pointing in any
direction independently of each other. Any typical configuration of
such a chain can be represented by the set of bond vectors
$\{\vec{a}_n\}$, $n=1,\ldots,N$, such that:
\begin{equation} \label{2.1}
\langle {\vec{a}_i}^2 \rangle = a^2=d\ell^2,
\end{equation}
here $d$ is the space dimension and the angular brackets stand for
averaging with respect to different possible orientations of each
bond. Fixing the starting point of the chain at the origin, one gets
for the end-to-end distance $\vec{R}_{\rm e}$ and its mean square:
\begin{equation} \label{2.2}
\vec{R}_{\rm e} = \sum_{i=1}^N \vec{a}_i\,, \qquad \langle
\vec{R}_{\rm e}^2 \rangle = N a^2.
\end{equation}
To get the second relation, one has to take into
account that $\langle \vec{a}_i \vec{a}_j \rangle = 0$  for $i\neq
j$, since random variables $\vec{a}_i$ are uncorrelated. Due to the
central limit theorem, the distribution function of the random
variable $R_{\rm e}=|\vec{R}_{\rm e}|$ takes on a Gaussian form:
\begin{equation} \label{2.3}
P(R_{\rm e}) =  \Big ( \frac{d}{2\pi \langle R_{\rm e}^2\rangle} \Big)^{d/2} \re^{-\frac{d
R_{\rm e}^2}{2\langle R_{\rm e}^2\rangle}}.
\end{equation}
Numerical factors in (\ref{2.3}) can be readily obtained from the
normalization conditions for the distribution function and its second
moment. Let us consider the three-dimensional ($d=3$) continuous
chain, such that
\begin{equation} \label{2.4}
P(R_{\rm e}){\rm d} R_{\rm e} =  \Big ( \frac{1}{2\pi N \ell^2} \Big)^{3/2}
\re^{-\frac{ R_{\rm e}^2}{2 N \ell^2 }} 4\pi R_{\rm e}^2 {\rm d} R_{\rm e}\,,
\end{equation}
defines the probability that the end point of the chain is located
in the interval between $R_{\rm e}$ and $R_{\rm e} + {\rm d}R_{\rm e}$\,. Following
Kuhn, let us take the mean value of $R_{\rm e}$ as one of the shape
characteristics of a chain and denote it by $\overline{r}_1\equiv
\langle R_{\rm e} \rangle$. One gets for its value:
\begin{equation} \label{2.5}
\overline{r}_1\equiv \int_0^\infty R_{\rm e} P(R_{\rm e}){\rm d} R_{\rm e} =  2\ell
\sqrt{\frac{2N}{\pi}}\,.
\end{equation}

To introduce two more observables that will characterize the shape of the chain, let us  proceed as follows. Having defined the end-to-end
vector $\vec{R_{\rm e}}$\,,  let us point the $z$-axis along this vector
(see figure~\ref{scheme}~(a)). Now, since both the starting and the end
points of the chain belong to the $z$-axis, the projection of the polymer coil on the
$xy$-plane has a form of a loop. It is natural to expect
that the largest deviation of a point on
this loop from the origin  corresponds to the $N/2$th step. Let us find the
distance in the plane from the origin to this point (we will denote it by $r_2$
hereafter) and take it as another shape characteristics of the
chain. Note, that vector $\vec{r}_2$ lies in the $xy$-plain and
therefore is two-dimensional. To do so, we have to find the distribution
function of a position vector of a point on a loop. It is convenient to find such a distribution
using the Bayes theorem for conditional probability~\cite{Bayes}. The theorem
relates the conditional and marginal probabilities of events $A$ and
$B$, provided that the probability of $B$ does not equal zero:
\begin{equation} \label{2.6}
P(A|B)=\frac{P(B|A)P(A)}{P(B)}\,.
\end{equation}
In our case, $P(A|B)\equiv P_2(r)$ is the probability that the
coordinate of the chain after $N/2$th step is given by a (now two-dimensional) vector $\vec{r}$ under the condition, that after $N$ steps
its coordinate is $r=0$. Then, $P(B|A)$ is the probability for the
chain that begins at the point with the coordinate $\vec{r}$ after
$N/2$ steps to return to the origin. Correspondingly, the prior
probability $P(A)$ is the probability that the coordinate of the
chain after $N/2$th step is given by a vector $\vec{r}$ (it is the
so-called ``unconditional'' or ``marginal'' probability of A, in our case it
is given by equation~(\ref{2.3}) for $d=2$). $P(B)$ is the prior or
marginal probability of B, in our case this is the probability for the
chain that starts at the origin to return back in $N$ steps, i.e. a
probability of a loop of $N$ steps. Realizing that for our case
$P(A)=P(B|A)$, that is probabilities to reach point $\vec{r}$
starting from  the origin is equal to the probability to reach the
origin starting from the point $\vec{r}$ we get:
\begin{equation} \label{2.7}
P_2(r)=P(A)^2/P(B),
\end{equation}
where $P(A)$ is given by equation~(\ref{2.3}) for $d=2$ and $P(B)$ may be
found from the normalization condition. In the continuous chain
representation we get for the probability that the $N/2$th bond of the
chain is located in the interval between $r$ and $r + {\rm d}r$:
\begin{equation} \label{2.8}
P_2(r){\rm d} r =  \frac{1}{2\pi N/2 \ell^2}\frac{1}{P(B)}
\re^{-\frac{ r^2}{2 N/2 \ell^2 }} \re^{-\frac{ r^2}{2 N/2 \ell^2 }} 2\pi
r {\rm d} r.
\end{equation}
The mean value $r_2$ follows:
\begin{equation} \label{2.9}
\overline{r}_2\equiv \int_0^\infty r P_2(r){\rm d} r = \frac{4}{N
\ell^2}\int_0^\infty \re^{-\frac{2\, r^2}{N \ell^2 }}  r^2  {\rm d} r
= \ell \sqrt{\frac{\pi N}{8}}\,.
\end{equation}

Now, let us point an $x$-axis along $\vec{r}_2$ (again see figure~\ref{scheme}~(a)) and repeat the above reasonings concerning the maximal
distance in $y$-coordinate, i.e. concerning the $y$-coordinate of the
$N/4$th step of the chain. We will denote it by $r_3$\,. The result
readily follows by the analogy with equation~(\ref{2.9}) taking into
account that now  the radius-vector is one-dimensional:
\begin{equation} \label{2.10}
\overline{r}_3\equiv \int_0^\infty r P_3(r){\rm d} r = \Big
(\frac{16}{\pi N \ell^2}\Big)^{1/2}\int_0^\infty \re^{-\frac{4\,
r^2}{N \ell^2 }} r  {\rm d} r = \frac{\ell}{2} \sqrt{\frac{N}{\pi}}\,.
\end{equation}

\begin{figure}[tb]
\begin{center}
\includegraphics[width=7cm]{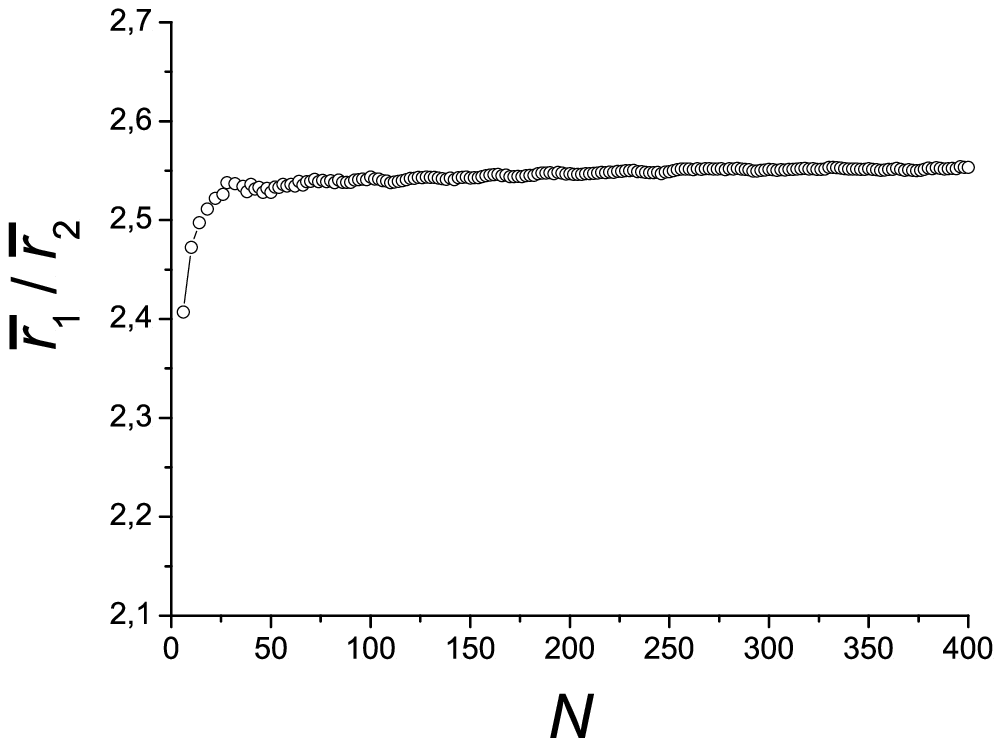}
\hfill
\includegraphics[width=7cm]{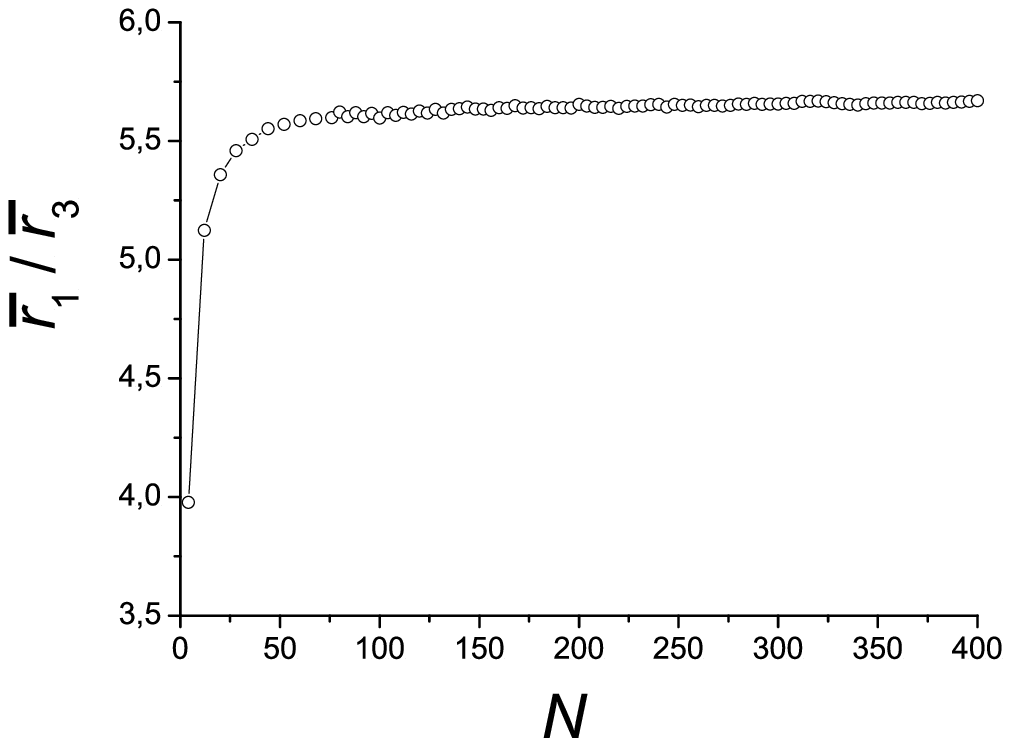}
\end{center}
\centerline{\hspace{3.8cm}{ a)} \hfill {  b)}\hspace{2.7cm} }
\caption{\label{r1r2r3} Ratios $\overline{r}_1/\overline{r}_2$ (a) and $\overline{r}_1/\overline{r}_3$ (b)
as functions of the chain length, results of computer simulations. Analytic estimates give:
$\overline{r}_1/\overline{r}_2 \simeq 2.55$ and $\overline{r}_1/\overline{r}_3 \simeq 5.66$
(see equations~(\ref{2.5}), (\ref{2.9}), (\ref{2.10})). }
\end{figure}

Comparing (\ref{2.5}), (\ref{2.9}), and (\ref{2.10}) one concludes that the
shape of the chain is characterized by three unequal sizes $\overline{r}_1$\,, $\overline{r}_2$\,,
and $\overline{r}_3$ with the following relations:
\begin{equation}\label{2.11}
\frac{\overline{r}_1}{\overline{r}_2}=\frac{8}{\pi}\simeq 2.55, \qquad
\frac{\overline{r}_1}{\overline{r}_3}=4\sqrt{2}\simeq 5.66.
\end{equation}
The above relations were first obtained by Kuhn~\cite{Kuhn34} and lead to the conclusion that a polymer chain even if considered as a chain of mutually intersecting steps (i.e. without account of an excluded volume effect) does not have a shape of a sphere but rather resembles an ellipsoid with unequal axes\footnote{As it was stated in Kuhn's paper,  the most probable shape of a polymer is a bend ellipsoid, of a bean-like shape: ``\dots verbogenes Ellipsoids (etwa die Form einer Bohne)\dots''~\cite{Kuhn34}.}.  To check how this prediction holds, we have performed numerical simulations of random walks on simple cubic lattices, constructing trajectories with the number of steps $N$ up to $400$ and performing the averaging over $10^6$ configurations.  As one can see from figure~\ref{r1r2r3}, the results of simulations are in perfect agreement with the data of~(\ref{2.11}).

Here it is worth mentioning  another far going prediction of the same paper~\cite{Kuhn34}. Discussing how might the excluded volume effect the polymer size, Kuhn arrives at the relation between the mean square end-to-end distance and the number of monomers which in our notations reads:
$\sqrt{\langle \vec{R}_{\rm e}^2 \rangle} = \ell N^{\nu}$ with $\nu=1/2+\epsilon$. Although this result is suggested on purely phenomenological grounds, its amazing feature is that the power-law form of the dependence is correctly predicted (cf. equation~(\ref{nu}) from the forthcoming section).  Moreover, Kuhn has estimated $\epsilon$ by considering the excluded volume effect for a $5$-segment chain for which he found an increase of $20\%$,
a result we have verified by exact enumerations. Assuming the power law form, $\epsilon$ is estimated as: $\epsilon\approx0.11$ and thus $\nu\approx0.61$, which is perfectly  confirmed later, e.g. by Flory theory~\cite{deGennes79}, which in $d=3$ gives $\nu=3/5$.  The notation for the correction used by Kuhn by coincidence is the same as that used much later in the famous $\varepsilon$-expansion~\cite{rgbooks} to develop a perturbation theory for calculation of this power law by means of the renormalization group technique (see equation~(\ref{nuSAW}) for the first order result). Before explaining how this theory is applied to calculate polymer shape characteristics, let us introduce observables in terms of which such description is performed.

\section{Description of polymer shape in terms of gyration tensor and combinations of its components\label{3}}

Let $\vec{R}_n=\{x_n^{1}\,,\ldots,x_n^d\}$ be the position vector of the $n$th monomer of a polymer chain ($n=1,\ldots,N$). The mean square of the end-to-end distance $R_{\rm e}$ of a chain thus reads:
\begin{equation}
\langle R_{\rm e}^2 \rangle =\langle |\vec{R}_N-\vec{R}_1|^2 \rangle \label{re},
\end{equation}
here and below, $\langle \ldots \rangle$ denotes the averaging over the ensemble of all
possible polymer chain configurations.
The basic shape properties of a specified spatial conformation of the chain can be characterized~\cite{Solc71,Rudnick86} in terms of the gyration tensor $\bf{Q}$ with components:
\begin{equation}
Q_{ij}=\frac{1}{N}\sum_{n=1}^N(x_n^i-{x^i_{\rm CM}})(x_n^j-{x^j_{\rm CM}}),\,\,\,\,\,\,i,j=1,\ldots,d,
\label{mom}
\end{equation}
with ${x^i_{\rm CM}}=\sum_{n=1}^Nx_n^i/N$ being the coordinates of the center-of-mass position vector ${\vec{R}_{\rm CM}}$\,.

\begin{figure}[h!]
\centerline{\includegraphics[width=4.9cm]{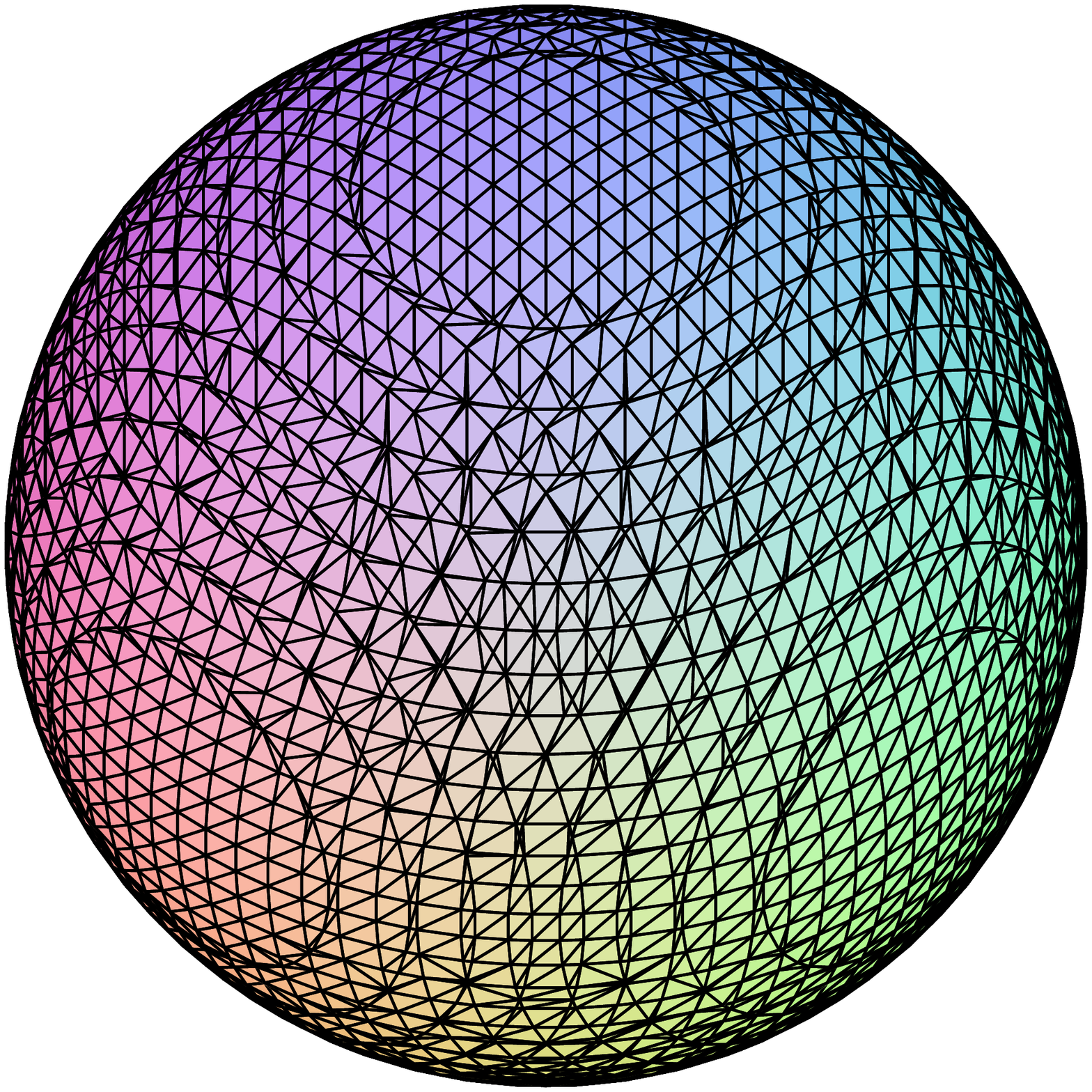}
\includegraphics[width=4.9cm]{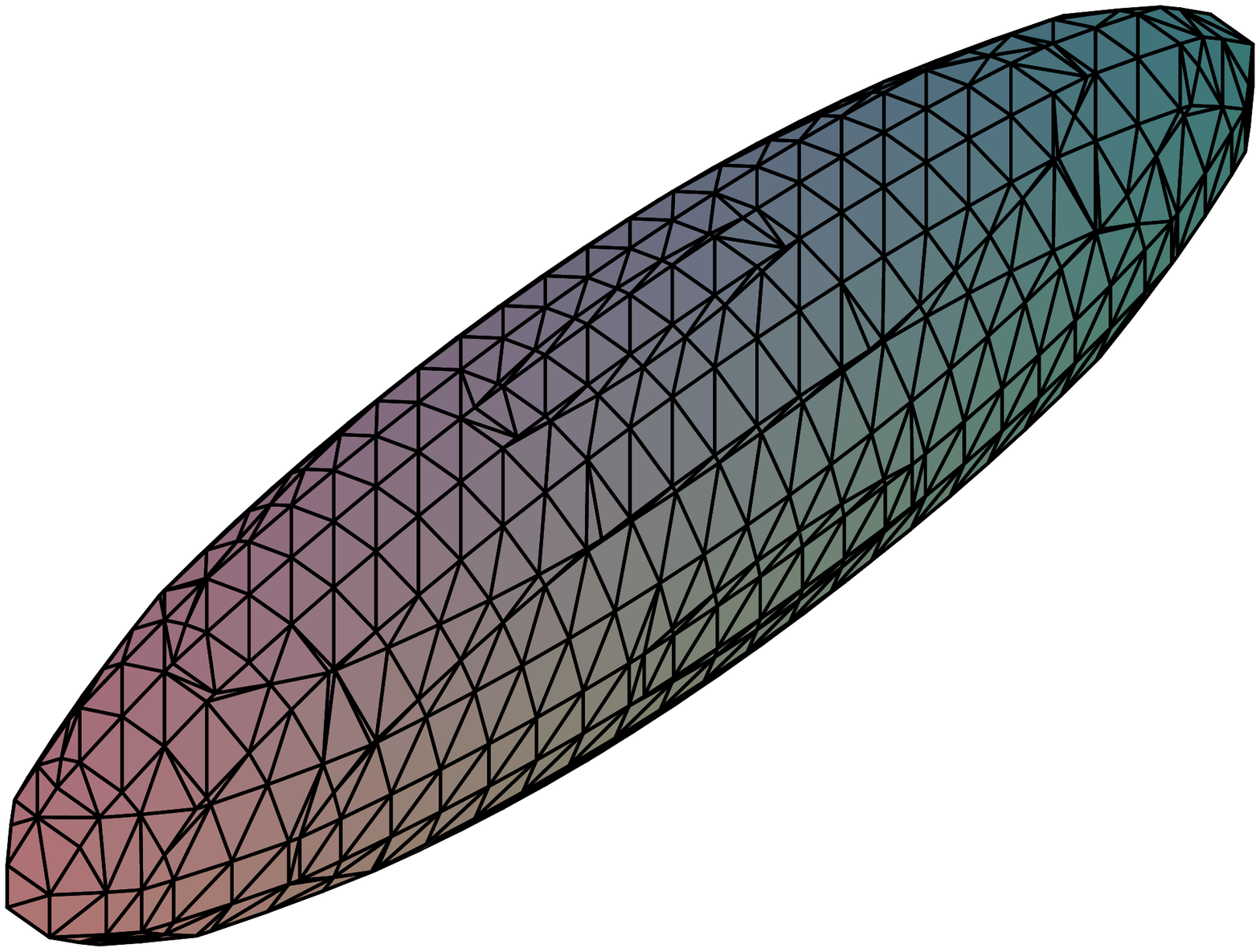}
\includegraphics[width=4.9cm]{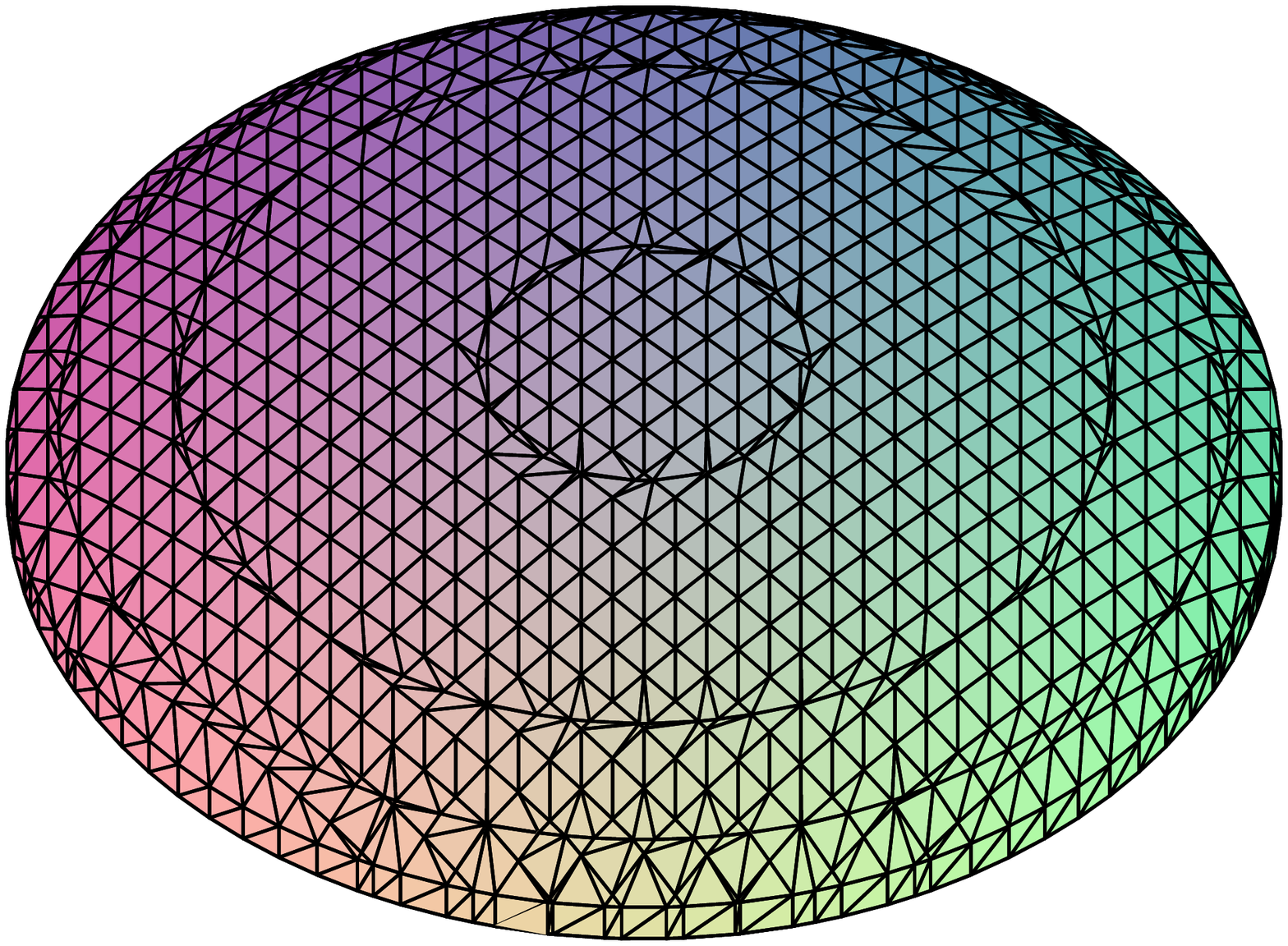}\\ \\}
\centerline{{  a)} \hspace{11em} {  b)} \hspace{11em} { c)}}
\caption{\label{fig1} Seen from far away, polymer coil may resemble the objects of different from. Here, we distinguish sphere-like {(a)}, prolate {(b)}, and oblate {(c)} conformations. Eigenvalues of corresponding gyration tensor (\ref{mom})
satisfy: $\lambda_1 \simeq \lambda_2 \simeq \lambda_3$
{(a)}, $\lambda_1 > \lambda_2 \simeq \lambda_3$ {(b)}, and $\lambda_1 \simeq \lambda_2 > \lambda_3$ {(c)}. Correspondingly, asphericity (\ref{add}) and prolateness (\ref{sdd}) of these conformations satisfy:
$A_d=0$, $S=0$ ({a}), $0< A_d < 1$, $0<S\leqslant 2$ ({b}), $0< A_d < 1$,
$-1/4\leqslant S <0$  ({c}). See the text for more details.}
\end{figure}

The spread in the eigenvalues $\lambda_i$ of the gyration tensor describes the distribution of monomers inside the polymer coil and
thus measures the asymmetry of the molecule; in particular, for a symmetric (spherical)
configuration all the eigenvalues $\lambda_{i}$ are equal, whereas for the so-called prolate  and oblate
configurations in $d=3$ (see figure~\ref{fig1}) the eigenvalues satisfy $\lambda_1\gg \lambda_2\approx\lambda_3$ and $\lambda_1\approx\lambda_2\gg\lambda_3$ correspondingly.
Solc and Stockmayer~\cite{Solc71} introduced the normalized average eigenvalues $\lambda_i$ of the gyration tensor as a shape measure of macromolecules.
Numerical simulations in $d=3$ dimensions give $\{\langle \lambda_1\rangle$,
 $\langle \lambda_2\rangle$, $ \langle \lambda_3\rangle \}$=$\{0.790,0.161,0.054 \}$~\cite {Kranbuehl77}, indicating a high anisotropy of typical polymer
configurations compared with the purely isotropic case $\{1/3$, $1/3$, $1/3 \}$.

While in simulations the eigenvalues of the gyration tensor ${\bf {Q}}$ can easily be calculated and averaged, different invariants have been devised for
theoretical calculations. As far as  ${\bf {Q}}$ has three  eigenvalues in $d=3$, one may construct three independent combinations of invariants. These are the square radius of gyration $R_{\rm G}^2$\,, the asphericity $A_d$ and the prolateness $S$ as elaborated in the following.
The first invariant of $\bf{Q}$ is the squared radius of gyration
\begin{equation}
R_{\rm G}^2 =\frac{1}{N} \sum_{n=1}^N (\vec{R}_n-{\vec{R}_{\rm CM}})^2  =  \sum_{i=1}^d Q_{ii} = {\rm Tr}\, \bf{Q}\,, \label{rg}
\end{equation}
which measures the distribution of monomers with respect to the center of mass.
To characterize the size measure of a single flexible polymer chain, one usually considers the mean-squared end-to-end distance $\langle R_{\rm e}^2 \rangle $ (\ref{re})
and radius of gyration
$\langle R_{\rm G}^2 \rangle$ (\ref{rg}), both
governed by the same scaling law:
\begin{equation}
 \langle R_{\rm e}^2 \rangle \sim \langle R_{\rm G}^2 \rangle \sim  N^{2\nu}, \label{nu}
\end{equation}
where  $N$ is the mass of the macromolecule (number of monomers in a polymer chain) and $\nu$ is a universal exponent ($\nu>1/2$ $(d<4)$, $\nu=1/2$ $(d\geqslant4)$).
The ratio of these two characteristic distances, the so-called size ratio:
\begin{equation}
g_d\equiv\langle R_{\rm e}^2 \rangle/\langle R_{\rm G}^2 \rangle\label{ggratio},
\end{equation}
 appears to be a universal, rotationally-invariant quantity ($g_d>6$ $(d<4)$, $g_d=6$ $(d\geqslant 4)$)~\cite{Witten78}.

Let ${\overline{\lambda}}\equiv {\rm Tr}\, {\bf{Q}}/d$
be the mean eigenvalue of  the gyration tensor.
Then one may characterize the extent of asphericity of a polymer chain configuration by the quantity ${A_d}$ defined as~\cite{Aronovitz86}:
\begin{equation}
{A_d} =\frac{1}{d(d-1)} \sum_{i=1}^d\frac{(\lambda_{i}-{\overline{\lambda}})^2}{\overline{\lambda}^2}=
\frac{d}{d-1}\frac{\rm {Tr}\,\bf{\hat{Q}}^2}{(\rm{Tr}\,{\bf{Q}})^2}\,, \label{add}
\end{equation}
with ${\bf{{\hat{Q}}}}\equiv{\bf{Q}}-\overline{\lambda}\,{\bf{I}}$ (here $\bf{I}$ is  the unity matrix).
This universal quantity equals zero for a spherical configuration, where all the eigenvalues are equal, $\lambda_i=\overline{\lambda}$, and takes a maximum value
of one in the case of a rod-like configuration, where all the eigenvalues equal zero except one. Thus, the inequality holds: $0\leqslant A_d\leqslant 1$.
Another rotational invariant quantity, defined in three dimensions, is the so-called prolateness $S$:
\begin{eqnarray}
S =\frac{\prod_{i=1}^3(\lambda_{i}-{\overline{\lambda}})}{{\overline{\lambda}}^3}=27\frac{{\rm det} \bf{{\hat{Q}}}}{(\rm{Tr}\,{\bf{Q}})^3}\,. \label{sdd}
\end{eqnarray}
If the polymer is absolutely prolate, rod-like  ($\lambda_1\neq0$, $\lambda_2=\lambda_3=0$), it is easy to see that $S$ equals two.
For absolutely oblate, disk-like conformations  ($\lambda_1=\lambda_2$\,, $\lambda_3=0$), this quantity takes on a value of $-1/4$,
while for a spherical configuration $S=0$.
In general, $S$ is positive for prolate ellipsoid-like polymer conformations  ($\lambda_1\gg \lambda_2\approx\lambda_3$) and negative for oblate
ones ($\lambda_1\approx\lambda_2\gg\lambda_3$), whereas its magnitude measures how much oblate or prolate the polymer is.
Note that since $\overline{\lambda}$ and the quantities in (\ref{rg})--(\ref{sdd}) are expressed in terms of rotational invariants, there is no need to explicitly determine the
eigenvalues $\lambda_i$ which greatly simplifies the calculations.

The average of the quantities (\ref{rg})--(\ref{sdd}) for a given polymer chain length $N$, denoted as $\langle \ldots \rangle$, is performed over an
ensemble of possible configurations of a chain.
 Note that some analytical and numerical approaches  avoid the averaging of the  ratio in (\ref{add}), (\ref{sdd})
and evaluate the ratio of averages:
\begin{equation}
\hat{A}_d=\frac{1}{d(d-1)}  \sum_{i=1}^d\frac{\langle(\lambda_{i}-{\overline{\lambda}})^2\rangle}{\langle\overline{\lambda}^2\rangle}\,, \qquad
\hat{S}= \frac{\prod_{i=1}^3\langle(\lambda_{i}-{\overline{\lambda}}) \rangle}{\langle{\overline{\lambda}}^3\rangle}\,, \label{avratio}
\end{equation}
which should be distinguished from the averaged asphericity and prolateness:
\begin{equation}
\langle{A_d}\rangle =\frac{1}{d(d-1)} \left \langle \sum_{i=1}^d\frac{(\lambda_{i}-{\overline{\lambda}})^2}{\overline{\lambda}^2} \right\rangle, \qquad
\langle{S}\rangle=\left\langle \frac{\prod_{i=1}^3(\lambda_{i}-{\overline{\lambda}})}{{\overline{\lambda}}^3} \right\rangle. \label{avad}
\end{equation}
Contrary to $\langle{A_d}\rangle$ and $\langle{S}\rangle$, the quantities~(\ref{avratio}) have no direct
relation to the probability distribution of the shape parameters $A_d$ and $S$.  As pointed out by Cannon~\cite{Cannon91}, this definition
overestimates the effect of larger polymer configurations on the mean shape properties and suppresses the effect of compact ones.
This artificially leads to overestimated values for shape parameters. The difference between $\langle{A_d}\rangle$ and $\hat{A}_d$  was found to be great
(see table~\ref{dani}).

\begin{table}[h!]
\caption{ \label{dani} Size ratio, averaged asphericity  and prolateness of flexible polymer chains on regular two- and three-dimensional lattices. MC: Monte Carlo simulations, DR: direct renormalization approaches.
 $a$: reference~\cite{Domb69}, $b$:~\cite{Bishop88}, $c$:~\cite{Benhamou85}, $d$:~\cite{Aronovitz86},
 $e$:~\cite{Jagodzinski92}.  }
%\begin{center}\begin{center}
\vspace{2ex}
\begin{center}
\begin{tabular}{lcccccc}
\hline
$d$ & Method & $g_d$ & $\langle A_d \rangle$& $\hat A_d$ & $\langle S \rangle$ & $\hat S$\\
\hline\\
2 & MC &  $7.14\pm0.03^a$ &  $0.501\pm 0.003^b$  &  $0.625\pm0.008^b$ & --  & --\\
2 & DR &  $7.003^c$ &  --        &  $0.558^d$  & --  & $0.899^d$\\
&&&&&\\
3 & MC & $6.249\pm0.03^e$ & $0.431\pm0.002^e$ & $0.546\pm0.008^b$ & $0.541\pm0.004^e$ &  -- \\
3 & DR &  $6.258^c$ & $0.415^e$  &   $0.529^d$  & --  & $0.893^d$\\
\hline
\end{tabular}\end{center}
%\end{center}
\end{table}

Numerous experimental studies indicate that a  typical flexible polymer chain in good solvent takes on the shape of an elongated, prolate ellipsoid, similar to what was shown in section \ref{3} for a random walk.
In particular,
using the data of x-ray crystallography and cryo-electron microscopy, it was found that the majority of non-globular proteins are characterized by $A_3$ values from $0.5$ to $0.7$
and $S$ values from $0$ to $0.6$~\cite{Dima04,Rawat09}. The shape parameters of polymers were analyzed analytically, based on the direct renormalization group approach~\cite{Aronovitz86,Jagodzinski92,Benhamou85}, and  estimated in numerical simulations~\cite{Solc71,Bishop88,Domb69,Honeycutt88}.
Table~\ref{dani} gives typical data for the above introduced shape characteristics of long flexible polymer chains in $d=2$ and $d=3$.

Since the above shape and size characteristics of polymer macromolecules are universal, i.e.
independent of the details of their  chemical structure, they were (along with polymer scaling exponents)
the subject of analysis by field-theoretical renormalization group approaches.
In the subsequent section we will introduce this approach as used to calculate  polymer shapes.

\section{Field-theoretical renormalization group approach to define \\ polymer shape \label{4}}

Aronovitz and Nelson~\cite{Aronovitz86} developed a scheme, allowing one to compute the universal shape parameters of
long flexible polymers within the frames of advanced field theory methods.

Here, we start with the Edwards continuous chain model~\cite{Edwards65},  representing
the polymer chain by a path ${\vec r}(t)$, parameterized by $0\leqslant t\leqslant T$ (see figure\ref{scheme} (b)).
The system can be described by the effective Hamiltonian ${\cal H}$:
\begin{eqnarray} \label{eff}
{\cal H}&=&\frac{1}{2}\int_0^{T}{\rm d}\,t
\left(\frac{{\rm d}\,{\vec r}(t)}{{\rm d} t}\right)^2{+}\frac{u_0}{4!}  \int_0^{T}{\rm d}
t\int_0^{T}{\rm d}t' \delta^d({\vec r}(t)-{\vec r}(t')).
\end{eqnarray}
The first term in (\ref{eff}) represents the chain connectivity, whereas
the second term describes the short range excluded volume interaction with coupling constant
$u_0$\,.

In this scheme, the gyration tensor components (equation~(\ref{mom}))  can be rewritten as:
\begin{equation}
\label{qab}
Q_{ij}=\frac{1}{2T^2}\int_0^T {\rm d}\,t_1 \int_0^T {\rm d}\,t_2\, \left[r^{i}(t_1)-r^{i}(t_2)\right]\,\left[r^{j}(t_1)-r^{j}(t_2)\right], \qquad i,j=1,\ldots,d.
\end{equation}

The model (\ref{eff}) may be mapped to a field theory
by a Laplace transform from the Gaussian surface
$T$ to the conjugated chemical potential variable (mass)
$\mu_0$ according to~\cite{Cloizeaux,Schafer91}:
\begin{equation} \label{laplace}
 \hat{\cal Z}(\mu_0)=\int{\rm d}T
\exp[-\mu_0 T]{\cal Z}(T),
\end{equation}
where ${\cal Z}(T)=\int D [ {\bf r} ]\exp(-{\cal H})$ is the partition  function of the system as function of the Gaussian surface and
$\int D [ {\bf r} ]$ means an integration over all possible path configurations~\cite{Kozitsky}.
Exploiting the analogy between the polymer problem and $O(m)$ symmetric field theory in the limit $m\to 0$
(de~Gennes limit)~\cite{deGennes79}, it was shown~\cite{Cloizeaux} that the partition function of the polymer system is related to the $m=0$-component field theory with an effective Lagrangean:
\begin{eqnarray}
{\cal L}&=&
\int{\rm d}^d
x\left[
 \frac{1}{2}(\mu_0^2
|\vec{\phi}(x)|^2{+}|\nabla\vec{\phi}(x)|^2
){+}\frac{u_0}{4!}(\vec{\phi}^2(x))^2
\right].\label{lag}
\end{eqnarray}
Here, $\vec{\phi}$ is an $m$-component vector field
$\vec{\phi}=(\phi^1,\ldots,\phi^m)$ and:
\begin{equation}
 \hat{\cal Z}(\mu_0)=\int D[\varphi] {\rm e}^{-{\cal L}}.
\end{equation}

One of the ways of extracting the scaling behavior of the model~(\ref{lag}),
is to apply the field-theoretical renormalization
group~(RG) method~\cite{rgbooks}  in the massive scheme, with the  Green's functions renormalized at
non-zero mass and zero external momenta. The Green's function
 $G_0^{(N,L)}$ can be defined as an average of $N$ field components and $L$ $\varphi^2$-insertions performed with the corresponding effective Lagrangean ${\cal L}$:
\begin{eqnarray}
\lefteqn{\delta(\sum k_i {+} \sum   p_j)
G_0^{(N,L)}(\{k\};\{p\}; \mu^2_0\,;u_0) =}\nonumber\\
&&\mbox{}\int^{\Lambda_0} \re^{\ri(k_i R_i{+}p_j r_j)}\langle\phi^2(r_1)\dots\phi^2(r_L) \phi(R_1)\dots \phi(R_N)
\rangle^{{\cal  L}_{\rm eff} }
{\rm d}^d R_1 \dots {\rm d}^d R_N {\rm d}^d r_1 \dots {\rm d}^d
r_L\,,
\end{eqnarray}
here $\{ k\}=(k_1\,,\ldots,k_L)$,  $\{p \}=(p_1\,,\ldots,p_N)$ are the sets of
external momenta and $\Lambda_0$ is a cut-off~\cite{30a}.
The renormalized Green's functions $G_{\rm R}^{(N,L)}$ are expressed in
terms of the bare vertex functions as follows:
\begin{eqnarray}
G_0^{(N,L)}(\{k\};\{p\};\hat\mu_0^2; u_0)=
Z_{\phi}^{N/2}Z_{\phi^2}^{-L}G_{\rm R}^{(N,L)}(\{k\};\{p\};\mu^2;u),\label{Green}
\end{eqnarray}
where
$Z_{\phi}$\,, $Z_{\phi^2}$ are the renormalizing factors, $\mu$,
$u$ are the renormalized mass and couplings.

The change of coupling constant $u_0$ under renormalization defines a flow in parametric space, governed by
the corresponding $\beta$-function:
\begin{equation}
\beta_u(u)=\frac{\partial u}{\partial \ln l}\Big|_0\,,\label{betau}
\end{equation}
where $l$ is the rescaling factor and $\big|_0$   stands for
 evaluation at fixed bare parameters.
 The fixed points (FP)
 of the RG transformation are given by the zeroes of the $\beta$-function.
The stable FP $u^*$, corresponding to the
critical point of the system, is defined as the fixed point where
$\frac{\partial \beta_u(u)}{\partial u}|_{u=u^*}
$
has a positive real part.
The flow of renormalizing factors $Z_{\phi}$\,, $Z_{\phi^2}$ defines the RG functions
$ \gamma_{\phi}(u)$, ${\bar \gamma}_{\phi^2}(u)$.
These functions, evaluated at the stable accessible FP, allow us to estimate the critical exponents.

Exploiting the perturbation theory expansion in parameter
$\varepsilon=4-d$ (deviation of space dimension from the upper critical one), one receives
within the  above described scheme up to the first order in $\varepsilon$ the well-known results for the fixed points:
\begin{eqnarray}
 u^*_{\rm RW}=0\,,\hspace{2.5mm} &   \text{stable for} &  \varepsilon\leqslant 0, \label{uRW}\\
 u^*_{\rm SAW}=\frac{3\varepsilon}{4}\,, & \text{stable for} &  \varepsilon>0.  \label{uSAW}
\end{eqnarray}
Here, $u_{\rm RW}$ describes the case of simple random walks (idealized polymer chain  without any intrachain interactions), and $u_{\rm SAW}$ is the fixed point, governing the scaling of self-avoiding random walks.

Evaluating the RG functions $ \gamma_{\phi}(u)$ and ${\bar \gamma}_{\phi^2}(u)$
at the above fixed points, one gets the familiar first-order results (see e.g.~\cite{rgbooks}) for the critical exponents $\nu$
and $\gamma$, that govern the scaling of the  polymer mean size (\ref{nu}) and the number of configurations, correspondingly:
\begin{eqnarray}
&&\nu_{\rm RW\phantom{I}}=\frac{1}{2}\,,  \qquad\quad\,\,\,  \gamma_{\rm RW\phantom{I}}=1\,, \label{nuRW}\\
&&\nu_{\rm SAW}=\frac{1}{2}{+}\frac{\varepsilon}{16}\,,  \quad\,\,\,  \gamma_{\rm SAW}=1{+}\frac{\varepsilon}{8}\,. \label{nuSAW}
%\label{nu_pure}
\end{eqnarray}

Following reference~\cite{Aronovitz86}, the averaged moments of gyration tensor $\bf{Q}$ (\ref{qab}), which are needed to determine
the polymer shape characteristics (\ref{rg}), (\ref{add}), (\ref{sdd}),
can be expressed in terms of renormalized connected Green's functions (\ref{Green}), in particular:
\begin{eqnarray}
&&\langle Q_{ij}\rangle=-\frac{1}{2}\left( \frac{T}{2\bar{X}}\right)^{2\nu}\frac{\Gamma(\gamma)}{\Gamma(\gamma{+}2\nu{+}2)}\frac{G_{ij}}{G_{\rm R}^{(2)}(0,0,u^*)}\,,
\label{qq}\\
&&\langle Q_{ij}Q_{kl}\rangle=-\frac{1}{4}\left( \frac{T}{2\bar{X}}\right)^{4\nu}
\frac{\Gamma(\gamma)}{\Gamma(\gamma{+}4\nu{+}4)}\frac{G_{ij|kl}}{G_{\rm R}^{(2)}(0,0,u^*)}\,.
\label{qqqq}
\end{eqnarray}
Here,  the following notations are used:
\begin{eqnarray}
&&G_{ij}\equiv \frac{\partial}{\partial q^{i}}\frac{\partial}{\partial q^{j}}\Big{|}_{q=0}G_{\rm R}^{(2,2)}(0,0;{q},-{ q};u^*), \label{Gii}\\
&&G_{ij|kl}={ \frac{\partial}{\partial q^{i}_1}\frac{\partial}{\partial q^{j}_1}\frac{\partial}{\partial q^{k}_2}\frac{\partial}{\partial q^{l}_2}\Big{|}_{{q}=0}G_{\rm R}^{(2,4)}(0,0;{ q}_1\,,-{ q}_1\,,{ q}_2\,,-{q}_2\,;u^*)},\label{Gij}
\end{eqnarray}
$\bar{X}$ is non-universal quantity, $\nu$ and $\gamma$ are the critical exponents,
${\partial}/{\partial q_{1}^i}$ means differentiation by the $i$-component of vector $q_1$\,,
$G_{\rm R}^{(2,2)}$ and $G_{\rm R}^{(2,4)}$
are the renormalized connected Green's function with 2 external legs and 2 insertions
$\varphi^2(q)$, $\varphi^2(-q)$ and 4 insertions $\varphi^2(q_1)$, $\varphi^2(-q_1)$,
$\varphi^2(q_2)$, $\varphi^2(-q_2)$ respectively, calculated at the fixed point for zero external momenta.

The isotropy of the original theory implies, in particular, that $\langle {\rm Tr {\bf {Q}}} \rangle =d\langle Q_{ii} \rangle$, so that:
\begin{equation}
\langle R_{\rm G}^2 \rangle = d \langle Q_{ii} \rangle.
\end{equation}
The mean-squared end-to-end distance $\langle R_{\rm e}^2 \rangle$ can be expressed as:
\begin{equation}
\langle R_{\rm e}^2 \rangle=-\left( \frac{T}{2\bar{X}}\right)^{2\nu}\frac{\Gamma(\gamma)}{\Gamma(\gamma{+}2\nu)}\frac{\left(\nabla_{{k}}^2G_{\rm R}^{(2)}({k},-{k},u^*)\right)\Big|_{{k}=0}}{G_{\rm R}^{(2)}(0,0,u^*)}\,.
\label{r}
\end{equation}
where $\nabla_{{k}}^2$ means differentiation over components of external momentum ${k}$.

One can easily convince oneself, that not-universal quantities cancel when
 the ratio~(\ref{ggratio}) is considered:
\begin{equation}
\label{ratio}
g\equiv \frac{\langle R_{\rm e}^2\rangle}{\langle R_{\rm G}^2\rangle} = \frac{\Gamma(\gamma{+}2\nu{+}2)}{\Gamma(\gamma{+}2\nu)}\frac{\left(\nabla_{{k}}^2G_{\rm R}^{(2)}(k,-k,u^*)\right)\Big|_{{k}=0}}
 {\left(\nabla_{{q}}^2G_{\rm R}^{(2,2)}(0,0;{q},-{q};u^*)\right)\Big|_{{q}=0}}\,.
\end{equation}
In derivation of~(\ref{ratio}) we made use of an
obvious relation $\sum_{i=1}^d G_{ii}=\nabla^2G_{\rm R}^{(2,2)}$.
 \begin{figure}[t!]
   \begin{center}
\includegraphics[width=8cm]{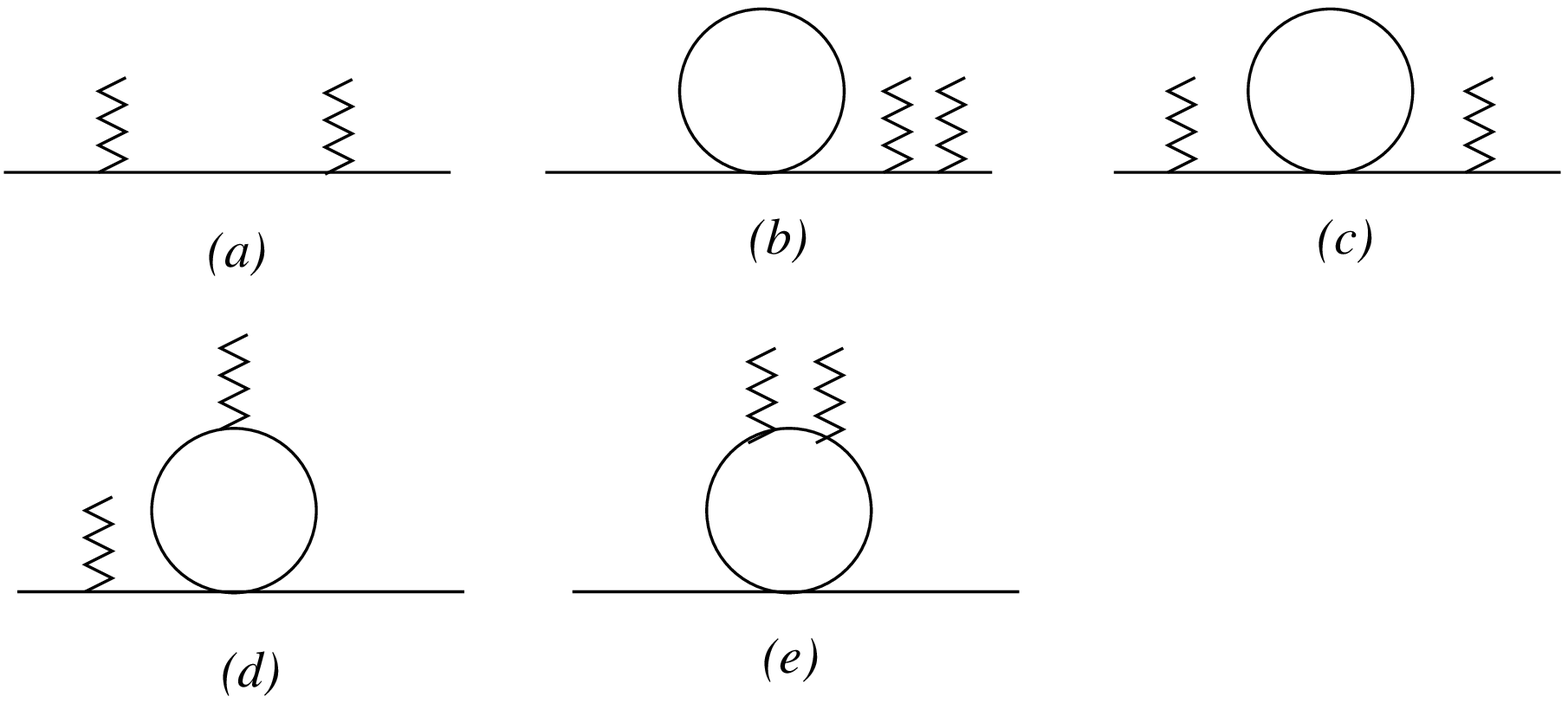}
\end{center}
\caption{\label{g22}Contributions to  the Green's function $G^{(2,2)}$ up to one-loop level. Solid lines denote propagators $\mu^2{+}k^2$, wavy lines illustrate the insertions of the type $\varphi^2$, loops imply integration over internal momenta. }
\end{figure}

\begin{figure}[t!]
\begin{center}
\includegraphics[width=8cm]{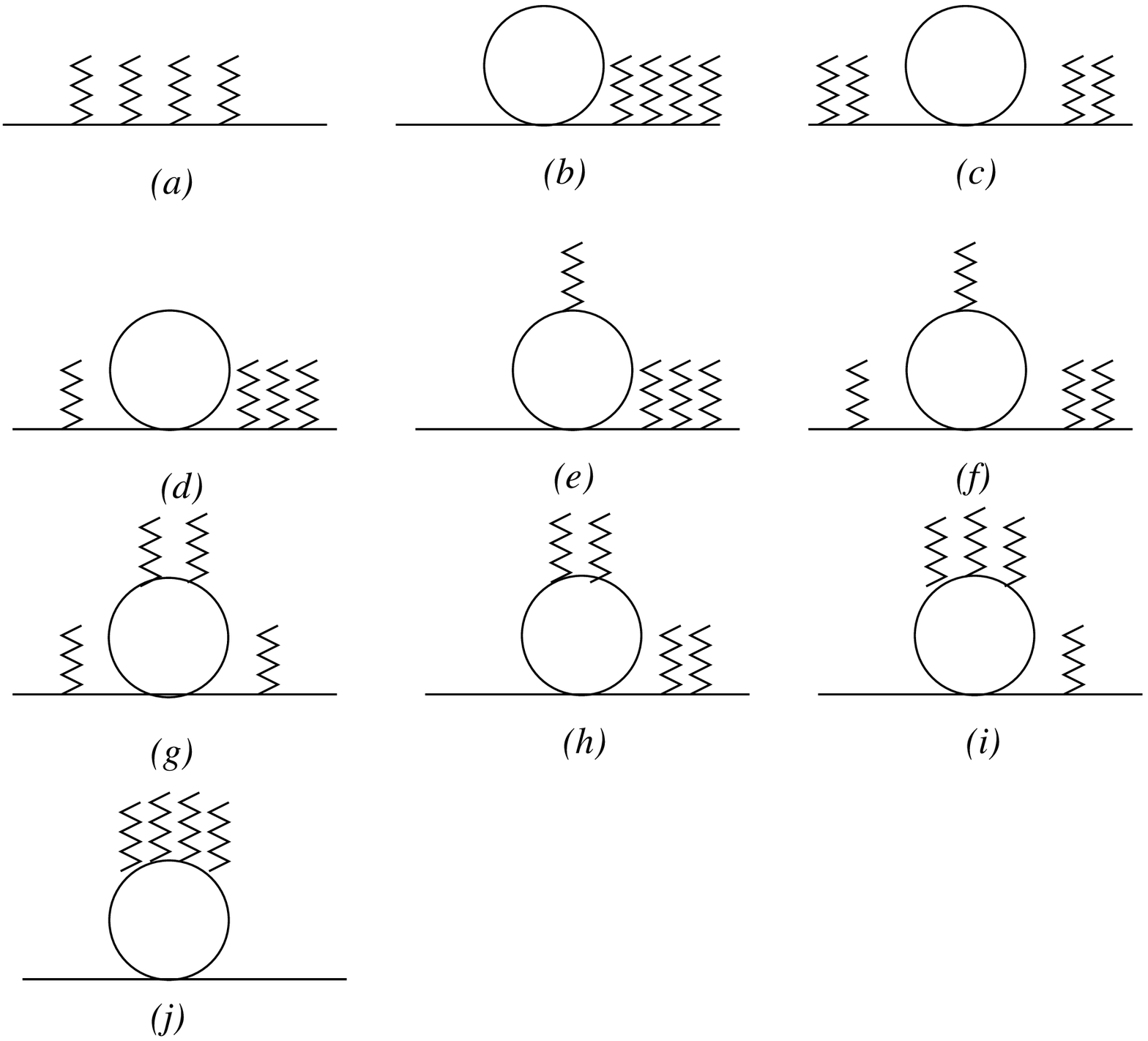}
\end{center}
\caption{\label{g24}Contributions to the Green's function $G^{(2,4)}$ up to one-loop level. Solid lines denote propagators $\mu^2{+}k^2$, wavy lines illustrate the insertions of the type $\varphi^2$, loops imply integration over internal momenta.}
\end{figure}

Computing the asphericity we follow equation~(\ref{avratio}) considering  $\hat A_d$ as the ratio of averages:
\begin{equation}
\hat A_d =\frac{d}{d-1}\frac{\langle{\rm {Tr}}\,{\bf{{\hat{Q}}}}^2\rangle}{\langle(\rm{Tr}\,{\bf{Q}})^2\rangle}\,. \label{ad1}
\end{equation}
This definition allows one to directly apply the renormalization group scheme described above,
and express the $ \hat A_d $ in terms of the averaged moments of the  gyration tensor (\ref{qq}), (\ref{qqqq}):
 \begin{equation}
\hat A_d  =\frac{\langle Q_{ii}^2\rangle{+}d\langle Q_{ij}^2\rangle-\langle Q_{ii}Q_{jj}\rangle}
{\langle Q_{ii}^2\rangle{+}d(d-1)\langle Q_{ii}Q_{jj}\rangle}\,,  \qquad i\neq j. \label{adfinal}
\end{equation}
One can again easily convince oneself that all the non-universal quantities in equations~(\ref{qq}),~(\ref{qqqq})
cancel when calculating~(\ref{adfinal}).
Note that within the RG approach we will focuss only on
two universal shape characteristics, namely the size ratio (\ref{ratio}) and asphericity (\ref{adfinal}), as far as the calculation of prolateness $S$ is particularly cumbersome.

To estimate~(\ref{ratio}), one needs to calculate
the Green's functions $G^{(2,2)}$, presented diagrammatically
in figure~\ref{g22}. Applying the  renormalization procedure, as described above,
one finds to the first order of the renormalized coupling $u$ (i.e., in the so-called one-loop approximation):
\begin{eqnarray}
&&G_{\rm R}^{(2,2)}(0,0;{q},-{q};u)=\frac{2}{q^2{+}1}-\frac{4}{3} \frac{1}{q^2{+}1}u I_1(0,q)
-\frac{2}{3}u I_2(0,0,q){+} \frac{4}{3}\frac{1}{q^2{+}1}u I_1(0,0),\label{g22pure}
\end{eqnarray}
the one-loop integrals $I_i$ are listed in appendix~A. Performing  an $\varepsilon=4-d$-expansion of
loop integrals (see appendix A for details) and differentiating over components of vector ${q}$,
 we found:
\begin{eqnarray}
\label{qq1}&&\left(\nabla_{{q}}^2G_{\rm R}^{(2,2)}(0,0;{q},-{q};u)\right)\Big|_{{q}=0}=-4-\frac{19}{18}u,\\
\label{kk2}&&\left( \nabla_{{k}}^2G_{\rm R}^{(2)}(k,-k)\right)\Big|_{{k}=0}=-2.
\end{eqnarray}
Substituting the values of fixed points (\ref{uRW}) and (\ref{uSAW}) into the ratio of these functions,
one receives the corresponding value of the $g$-ratio:
\begin{eqnarray}
&& g_{\rm RW\phantom{I}} = 6\,, \label{gratio_RW}\\
&& g_{\rm SAW} = 6{+}\frac{\varepsilon}{16}\,.\label{gratio_pure}
\end{eqnarray}
Equation~(\ref{gratio_pure}) presents a first order correction to the $g$-ratio caused by excluded volume interactions.

To compute the averaged asphericity ratio using~(\ref{adfinal}), one needs  the Green
 function $G^{(2,4)}$ with four insertions $\varphi^2({q}_1)/2$, $\varphi^2(-{q}_1)/2$, $\varphi^2({q}_2)/2$, $\varphi^2(-{q}_2)/2$, which is schematically presented in figure~\ref{g24}.
Applying the renormalization scheme described above, one receives an analytic expression for the renormalized function $G_{\rm R}^{(2,4)}$, given in the appendix~B, equation~(\ref{g24pure}).
Taking derivatives over components of inserted vectors ${q}_1$\,, ${q}_2$ and performing $\varepsilon$-expansions of the resulting expressions (see appendix~A for details) we arrive at the following expansions for the functions~(\ref{Gii}),~(\ref{Gij}):
\begin{eqnarray}
&&G_{xx\phantom{|yy}}=576{+}\frac{4028 u}{15}\,,\nonumber\\
&&G_{xx|yy}=320{+}\frac{436 u}{3}\,,\nonumber\\
&&G_{xy|xy}=128{+}\frac{308 u}{5}\,.
\end{eqnarray}
Evaluating these relations at the fixed points (\ref{uRW}), (\ref{uSAW})  and substituting into (\ref{qq}), (\ref{qqqq}), (\ref{adfinal}) one receives:
 \begin{eqnarray}
&& \hat A_d^{\rm RW\phantom{I}}=\frac{1}{2}\,,\label{ad_RW}\\
&& \hat A_d^{\rm SAW}=\frac{1}{2}{+}\frac{15}{512}\varepsilon\,.\label{ad_SAW}
\end{eqnarray}
We note, that result (\ref{ad_RW}) means that even within the idealized model of simple random walks, the shape of a polymer chain is
highly anisotropic (cf. Kuhn's picture, described in section 2). Taking into account the excluded volume effect makes the polymer chain more extended and  aspherical: indeed, in three dimensions ($\varepsilon=1$) the above obtained quantity reads: $\hat A_d^{\rm SAW}\simeq0.53$.

\section{Polymer in porous environment: Model with long-range correlated disorder \label{5}}

In real physical processes, one is often interested how structural obstacles (impurities) in the environment
alter the behavior of a system. The density fluctuations of obstacles lead to a large spatial inhomogeneity and create pore spaces,
which are often of fractal structure~\cite{Dullen79}.
In polymer physics, it is of great importance to understand the behavior of macromolecules in the presence of structural disorder,
e.g., in colloidal solutions~\cite{Pusey86} or near the microporous membranes~\cite{Cannel80}.
In particular, a related problem concerns the protein folding dynamics in the cellular environment,
which can be considered as a highly disordered environment due to the presence of a large amount of biochemical species, occupying up to $40\%$ of the total volume~\cite{Minton01}.
Structural obstacles strongly affect the protein folding~\cite{protein}.
Recently, it was realized experimentally~\cite{Samiotakis09} that macromolecular crowding has a dramatic effect on the shape properties of proteins.

In the language of lattice models, a disordered environment with structural obstacles can be considered as a lattice, where some amount
of randomly chosen sites contain defects which are to be avoided by the polymer chain.
Of particular interest is the case when the concentration of lattice sites allowed for the SAWs
equals the critical concentration and the lattice is at the percolation threshold.
In this regime, SAWs belong to a new universality class, the scaling law~(\ref{nu}) holds with exponent $\nu_{\rm p_c}>\nu_{{\rm SAW}}$~\cite{sawperc}.
The universal shape characteristics of
flexible polymers in disordered environments modeled by a percolating lattice were studied recently in~\cite{Janke10}.
Another interesting situation arises when the structural obstacles of environment display
correlations on a mesoscopic scale~\cite{Sahimi95}.
One can describe such a medium by a model of long-range-correlated
(extended) quenched defects. This model was proposed
in reference~\cite{Weinrib83} in the context of magnetic phase
transitions.  It considers defects, characterized by a pair
correlation function $h(r)$, that decays with a distance $r$
according to a power law:
\begin{equation}\label{paircor}
h(r)\sim r^{-a}
\end{equation}
at large $r$.
This type of disorder has a direct interpretation
for integer values of $a$; namely, the case $a=d$ corresponds to
point-like defects, while $a=d-1$ $(a=d-2)$ describes straight lines
(planes) of impurities of random orientation. Non-integer values of $a$
are interpreted in terms of impurities organized in fractal
structures~\cite{fractals}. The effect of this type of disorder
on the magnetic phase transitions has been  a subject to numerous studies~\cite{lr_magn}.

The impact of long-range-correlated disorder on the scaling of single polymer chains was analyzed in
our previous works~\cite{Blavatska0102} by means of field-theoretical renormalization group approach. In particular, it was shown
 that the correlated obstacles in environment lead to a new
universality class with values of the polymer scaling exponents
that depend on the strength of correlation expressed by parameters $a$. The question about how the characteristics of shape of a
flexible chain are effected by the presence of such a porous medium was briefly discussed  by us in reference~\cite{Blavatska10}.
The details of these calculations will be presented here.

We introduce disorder into the model (\ref{lag}), by redefining
$\mu_0^2 \to \mu_0^2{+}\delta\mu_0(x)$,  where the
 local fluctuations
$\delta\mu_0(x)$ obey:
\begin{align}
 \langle\langle\delta\mu_0(x)\rangle\rangle&=0,\nonumber\\
\langle\langle\delta\mu_0(x)\delta\hat\mu_0(y)\rangle\rangle \label{corre}
&=h(|x-y|).
\end{align}
Here, $\langle\langle\ldots\rangle\rangle$ denotes
the average over spatially homogeneous and isotropic quenched
disorder. The form of the pair correlation function $h(r)$ is
chosen to decay with distance according to~(\ref{paircor}).

In order to average the free energy over different configurations
of the quenched disorder we apply the replica method to construct
an effective Lagrangean~\cite{Blavatska0102}:
\begin{eqnarray}
 {\cal L}_{\rm dis}&=& \frac{1}{2} \sum_{\alpha=1}^{n}
\int{\rm d}^d x \left[\left(\mu_0^2|\vec{\phi}_{\alpha}(x)|^2{+}
|\nabla\vec{\phi}_{\alpha}(x)|^2\right)
{+}\frac{u_0}{4!}\left(\vec{\phi}^2_{\alpha}(x)\right)^2\right]
\nonumber\\
&{+}&\sum_{\alpha,\beta=1}^{n}
\int{\rm d}^dx{\rm d}^dy h(|x-y|)
\vec{\phi}_{\alpha}^2(x)\vec{\phi}_{\beta}^2(y).
\label{Leff}
\end{eqnarray}
Here, the term describing replicas coupling contains the correlation
function  $h(r)$ (\ref{paircor}), Greek indices denote
replicas and both the replica ($n\to 0$) and polymer ($m\to 0$)
limits are implied.
For small $k$, the Fourier-transform $\tilde h(k)$ of (\ref{paircor}) reads:
\begin{equation}
\tilde h(k)\sim v_0{+}w_0|k|^{a-d}.
\label{gkk}
\end{equation}
Taking this into account, rewriting equation~(\ref{Leff}) in  momentum
space variables, and recalling the special symmetry properties of~(\ref{Leff}) that appear for $m$, $n\to 0$~\cite{Blavatska0102},  a theory with
two bare couplings $u_0$\,, $w_0$ results.
Note that for $a\geqslant d$ the $w_0$-term is irrelevant in
the RG sense and one restores the pure case (absence of structural disorder). As it will be shown below, this term modifies the critical behaviour at $a<d$. We will refer to this type of disorder as long-range-correlated and denote by LR hereafter.

To extract the scaling behavior of the model~(\ref{Leff}),
one applies the field-theoretical renormalization
group  method following the scheme described in a previous section, with modifications caused by the presence of the second coupling constant $w_0$\,. In particular,
the change of couplings $u_0$\,, $w_0$ under renormalization defines a flow in parametric space, governed by
corresponding $\beta$-functions (c.f. equation (\ref{betau})):
\begin{equation}
\beta_u(u,w)=\frac{\partial u}{\partial \ln l}\Big|_0\,, \qquad
\beta_w(u,w)=\frac{\partial w}{\partial \ln l}\Big|_0\,.\label{betafunc}
\end{equation}
 The fixed points (FPs)
 of the RG transformation are given by common zeroes of the $\beta$-functions.
The stable FP ($u^*$, $w^*$) that corresponds to the
critical point of the system, is defined as the fixed point where
the stability matrix
$B_{ij}={\partial \beta_{\lambda_i}}/{\partial \lambda_j}$\,, $i,j=1,2$
possesses eigenvalues with positive real parts (here, $\lambda_1=u$, $\lambda_2=w$).

In our previous work~\cite{Blavatska0102} we have found the FP
coordinates for polymers in LR disorder, which up to the first order of  $\varepsilon=4-d$, $\delta=4-a$-expansion read:
\begin{align}
&u^*_{\rm RW\phantom{I}}=0, & &w^*_{\rm RW\phantom{I}}=0 & &\text{stable for} \quad \delta<0,\, \delta<0,\label{RW}\\
&u^*_{\rm SAW}=\frac{3\varepsilon}{4}\,, & &w^*_{\rm SAW}=0 & &\text{stable for} \quad \delta<\varepsilon/2,\label{pure}\\
&u^*_{\rm LR\phantom{B}}=\frac{3\delta^2}{2(\varepsilon-\delta)}\,, & &w^*_{\rm LR \phantom{B}}=\frac{3\delta(\varepsilon-2\delta)}{2(\varepsilon-\delta)} & &\text{stable for} \quad \varepsilon/2 <\delta<\varepsilon. \label{uLR}
\end{align}
The RW and SAW fixed points restore the corresponding cases of a polymer in pure solvent
(cf.~(\ref{uRW}),~(\ref{uSAW})),
whereas the LR fixed point reflects the effect of correlated obstacles, which appears to be non-trivial in certain regions of the $d$, $a$ plane and to govern  new scaling behaviour in this region.
For critical exponents $\nu_{\rm LR}$, $\gamma_{\rm LR}$ governing the scaling of polymer chains in the region of $a$, $d$,
where the effect of LR disorder is nontrivial, one obtains~\cite{Blavatska0102}:
\begin{equation}
\nu_{\rm LR}= 1/2 {+} \delta/8, \qquad \gamma_{\rm LR}=1{+}\delta/4. \label{nuLR}
\end{equation}

As it was explained in the previous section, to estimate the universal size ratio (\ref{ratio}) for the case of polymers in long-range-correlated disorder, we calculate
the Green function $G^{(2,2)}(u,w)$, presented diagrammatically in figure~\ref{g22}. Now, every interactive diagram appears twice, once with each of the two  couplings $u$ and $w$, respectively.  For the renormalized function we obtain up to the  one-loop approximation:
\begin{eqnarray}
&&G_{\rm R}^{(2,2)}(0,0;{q},-{q};u,w)=\frac{2}{q^2{+}1}-\frac{4}{3} \frac{1}{q^2{+}1}\left[u I_1(0,q)-w J_1(0,q)\right]\nonumber\\
&&-\frac{2}{3}\left[ u I_2(0,0,q)-w J_2(0,0,q)\right]{+} \frac{4}{3}\frac{1}{q^2{+}1}
\left[ u I_1(0,0)-w J_1(0,0)\right].\label{g22LR}
\end{eqnarray}
The one-loop integrals  $I_i$ are given in appendix~A. Note that $J_i$ differs from corresponding $I_i$  only by an
additional factor $|p|^{a-d}$ in the numerator.
Differentiating over components of vector ${q}$, evaluating at ${q}=0$ and performing double $\varepsilon,\delta$-expansions of the  resulting expression, we have:
\begin{eqnarray}
\label{q1}&&\left(\nabla_{{q}}^2G_{\rm R}^{(2,2)}(0,0;{q},-{q};u,w)\right)\Big|_{{q}=0}=-4-\frac{19}{18}u-\frac{19}{18}w,\\
\label{k2}&&\left(\nabla_{{k}}^2G_{\rm R}^{(2)}(k,-k)\right)\Big|_{{k}=0}=-2.
\end{eqnarray}
Recalling the value of the LR fixed point (\ref{uLR}),
and inserting this into the ratio (\ref{ratio}) of these functions,  one arrives at an expansion for the size ratio of a SAW in the presence of long-range correlated disorder:
\begin{equation}
g_{\rm LR}= 6{+}\frac{\delta}{8}\,. \label{gratio_LR}
\end{equation}
This should be compared with the corresponding value in the pure case (\ref{gratio_pure}). Let us qualitatively estimate the change in the size ratio $ g$, caused by the presence of structural obstacles in three dimensions. Substituting directly $\varepsilon=1$ into (\ref{gratio_pure}), we have for the
polymer chain in a pure solvent:
$ g^{{\rm pure}}\simeq 6.06. $  Let us recall that the effect of long-range-correlated disorder is relevant to $a\leqslant d$ ($\delta\geqslant\varepsilon$) (see e.g. explanation after equation~(\ref{Leff})).
Estimates of $ g_{{\rm LR}}$ can be evaluated by direct substitution of the continuously variable parameter
$\delta$ into equation~(\ref{gratio_LR}).
One concludes that increasing the parameter $\delta$ (which corresponds to an increase of
disorder strength) leads to a corresponding increase of the $g$-ratio.

To compute the averaged asphericity in correspondence with (\ref{adfinal}), we need the Green function $G^{(2,4)}(u,w)$ with four insertions $\varphi^2({q}_1)/2$, $\varphi^2(-{q}_1)/2$, $\varphi^2({q}_2)/2$, $\varphi^2(-{q}_2)/2$, shown diagrammatically
in figure~\ref{g24}. The corresponding analytic expression for the renormalized function is given in appendix~B, equation~(\ref{g24LR}).
Taking derivatives of this expression with respect to the components of the inserted vectors ${q}_1$, ${q}_2$
we find:
\begin{eqnarray}
&&G_{xx\phantom{|yy}}=576{+}\frac{4028}{15}(u-w),\nonumber\\
&&G_{xx|yy}=320{+}\frac{436}{3}(u-w),\nonumber\\
&&G_{xy|xy}=128{+}\frac{308}{5}(u-w).
\end{eqnarray}
At the LR fixed point (\ref{uLR}), from (\ref{adfinal}) we finally have:
\begin{equation}
\hat A_d^{\rm LR}= \frac{1}{2}{+}\frac{1}{48}\,\varepsilon{+}\frac{13}{768} \,\delta.\label{adLR}
\end{equation}
Again, let us qualitatively estimate the change in $\hat A_d $  caused by
the presence of structural obstacles in three dimensions. Substituting directly $\varepsilon=1$ into~(\ref{ad_SAW}), we have for the pure case: $ \hat{A}_d^{{\rm pure}}\simeq 0.53$. Estimates of $\hat{A}_d^{{\rm LR}}$ can be obtained by direct
substitution of the continuously changing parameter $\delta$ into  equation~(\ref{adLR}).
An increasing strength of disorder correlations results in
an increase of the asphericity  ratio of polymers in disorder.
This phenomenon is intuitively understandable if one recalls the impact
of the long-range-correlated disorder on the mean end-to-end distance
exponent~$\nu$. Indeed, it has been shown
in~\cite{Blavatska0102} that such disorder leads to
an increase of~$\nu$, and subsequently, to the swelling of a polymer chain.
Extended obstacles do not favour return trajectories and as a result the polymer chain
becomes more elongated. In turn, such elongation leads to an increase of the
asphericity ratio as predicted by equation~(\ref{adLR}).

\section{Conclusions \label{7}}

The universal characteristics of the average shape of polymer coil configurations in a
porous (crowded) environment with structural obstacles have been analyzed considering the special case
when the defects are correlated  at large distances $r$ according to a power law: $h(r)\sim r^{-a}$.
Applying the field-theoretical RG approach, we estimate the size
 ratio $ g=\langle R_{\rm e}^2 \rangle/\langle R_{\rm G}^2 \rangle$ and averaged asphericity ratio $\hat{A_d}$ up
 to the first
 order of a double $\varepsilon=4-d$, $\delta=4-a$ expansion. We have revealed that
 the presence of long-range-correlated disorder leads to an increase of both $g$ and $\hat{A_d}$
 as compared to their values for a polymer chain in a pure solution. Moreover,
 the value of the asphericity ratio $\hat{A_d}$  was found to be closer
to the maximal value of one in presence of correlated obstacles.
Thus, we  conclude that the presence of structural obstacles in an environment
enforces the polymer coil configurations to be less spherical. We believe that the obtained first order
results indicate the appropriate qualitative  changes in polymer shape caused by long-range-correlated
environment. However,  to get more accurate quantitative results, higher orders of perturbation theory may be
needed. This is  subject of further investigations.

\section*{Acknowledgements}
We thank Prof. Myroslav Holovko (Lviv) for an
invitation to submit a paper to this Festschrift and Prof. Yuri Kozitsky (Lublin)
for useful discussions.
This work was supported in part by the
FP7 EU IRSES project N269139  ``Dynamics and Cooperative Phenomena in Complex
Physical and Biological Media'' and Applied Research Fellowship of Coventry University.

\appendix

\section*{Appendix A \label{A}}
\renewcommand{\theequation}{A\arabic{equation}}

Here, we present the expressions for the loop integrals, as they appear
in the Green functions $G^{(2,2)}$ and $G^{(2,4)}$. We make the couplings dimensionless by
redefining $u=u\,\mu^{d-4}$ and
$w=w\,\mu^{a-4}$; therefore, the loop integrals do not explicitly
contain the mass~\cite{30a}:
\begin{eqnarray}
&&I_1(k_1, k_2)=\int\frac{{\rm d}{\vec p}}{\left[(p{+}k_1)^2{+}1\right]\left[(p{+}k_2)^2{+}1\right]}\,,\nonumber\\
&&I_2(k_1, k_2\,, k_3)=\int\frac{{\rm d}{\vec p}}{[(p{+}k_1)^2{+}1][(p{+}k_2)^2{+}1][(p{+}k_3)^2{+}1]}\,,\nonumber\\
&&I_3(k_1, k_2\,,k_3\,,k_4)=\int\frac{{\rm d}{\vec p}}{[(p{+}k_1)^2{+}1][(p{+}k_2)^2{+}1][(p{+}k_3)^2{+}1][(p{+}k_4)^2{+}1]}\,,\nonumber\\
&&I_4(k_1, k_2\,, k_3\,, k_4\,, k_5)=\int\frac{{\rm d}{\vec p}}
{[(p{+}k_1)^2{+}1][(p{+}k_2)^2{+}1][(p{+}k_3)^2{+}1][(p{+}k_4)^2{+}1][(p{+}k_5)^2{+}1]}\,.
\nonumber
\end{eqnarray}
The loop integrals $J_i$ (equations~(\ref{g22LR}) and~(\ref{g24LR})) differ from corresponding $I_i$  only by
an additional factor $|p|^{a-d}$ in the numerators.
The correspondence of the integrals to the diagrams in
figure~\ref{g24} is:
\begin{eqnarray*}
(e),  (f): &&\text{integrals} \quad I_1(0, q_1), \, I_1(0, q_2), \, J_1(0, q_1), \, J_1(0, q_2), \\
(g): &&I_2(0, q_1\,, q_2), \, I_2(0, 0, q_1), \, I_2(0, 0, q_2), \,
J_2(0, q_1\,, q_2), \, J_2(0, 0, q_1), \, J_2(0, 0, q_2),\\
&&J_2(0, q_1\,, q_2), \, J_2(0, 0, q_1), \, J_2(0, 0, q_2),\\
(h): &&I_2(0, 0, q_1), \, I_2(0, 0, q_2), \, I_2(0, q_1\,, q_1{+}q_2), \, I_2(0, q_2\,, q_1{+}q_2), \, J_2(0, 0, q_1),\\
&&J_2(0, 0, q_2), \, J_2(0, q_1\,, q_1{+}q_2), \, J_2(0, q_2\,, q_1{+}q_2),\\
\end{eqnarray*}
%\vspace{-1cm}
\begin{eqnarray*}
(i): &&I_3(0, 0, q_1\,, q_2), \, I_3(0, q_1\,, q_2\,, q_1{+}q_2), \, I_3(0, q_1\,, q_1\,, q_1{+}q_2), \, I_3(0, q_2\,, q_2\,, q_1{+}q_2), \\
&&J_3(0,  0, q_1\,, q_2), \, J_3(0,  q_1\,,  q_2\,,  q_1{+}q_2), \, J_3(0,  q_1\,,  q_1\,,  q_1{+}q_2), \, J_3(0, q_2\,, q_2\,, q_1{+}q_2), \\
(j): &&I_4(0, 0, 0, q_1\,, q_2), \, I_4(0, 0, q_1\,, q_2\,, q_1{+}q_2), \, I_4(0, 0, q_1\,, q_1\,, q_1{+}q_2), \, I_4(0, 0, q_2\,, q_2\,, q_1{+}q_2), \\
&&J_4(0, 0, 0, q_1\,, q_2), \, J_4(0, 0, q_1\,, q_2\,, q_1{+}q_2), \, J_4(0, 0, q_1\,, q_1\,, q_1{+}q_2), \, J_4(0, 0, q_2\,, q_2\,, q_1{+}q_2).
\end{eqnarray*}
%$(e)$,  $(f)$: integrals $I_1(0, q_1)$,  $I_1(0, q_2)$,  $J_1(0, q_1)$,  $J_1(0, q_2)$, \\
%$(g)$: $I_2(0, q_1, q_2)$,  $I_2(0, 0, q_1)$,  $I_2(0, 0, q_2)$,
%$J_2(0, q_1, q_2)$,  $J_2(0, 0, q_1)$,  $J_2(0, 0, q_2)$\\
%$(h)$: $I_2(0, 0, q_1)$,  $I_2(0, 0, q_2)$,  $I_2(0, q_1, q_1{+}q_2)$,  $I_2(0, q_2, q_1{+}q_2)$,
%$J_2(0, 0, q_1)$,  $J_2(0, 0, q_2)$,  $J_2(0, q_1, q_1{+}q_2)$,  $J_2(0, q_2, q_1{+}q_2)$\\
%$(i)$: $I_3(0, 0, q_1, q_2)$,  $I_3(0, q_1, q_2, q_1{+}q_2)$,  $I_3(0, q_1, q_1, q_1{+}q_2)$,  $I_3(0, q_2, q_2, q_1{+}q_2)$,
%$J_3(0,  0, q_1, q_2)$,\\  $J_3(0,  q_1,  q_2,  q_1{+}q_2)$,  $J_3(0,  q_1,  q_1,  q_1{+}q_2)$,  $J_3(0, q_2, q_2, q_1{+}q_2)$, \\
%$(j)$: $I_4(0, 0, 0, q_1, q_2)$,  $I_4(0, 0, q_1, q_2, q_1{+}q_2)$,  $I_4(0, 0, q_1, q_1, q_1{+}q_2)$,  $I_4(0, 0, q_2, q_2, q_1{+}q_2)$, \\
%$J_4(0, 0, 0, q_1, q_2)$,  $J_4(0, 0, q_1, q_2, q_1{+}q_2)$,  $J_4(0, 0, q_1, q_1, q_1{+}q_2)$,  $J_4(0, 0, q_2, q_2, q_1{+}q_2)$.\\

In our calculations,  we use the following formula to fold many
denominators into one (see e.g. book of D.~Amit in  reference~\cite{rgbooks}):
%\begin{widetext}
\begin{eqnarray}
%\lefteqn{
\lefteqn{\frac{1}{a_1^{\alpha_1}\ldots a_n^{\alpha_n}}=\frac{\Gamma(\alpha_1{{+}}\ldots{{+}}\alpha_n)}
{\Gamma(\alpha_1)\ldots
\Gamma(\alpha_n)}\times} \nonumber\\
&&\mbox{}\times\int\limits_0^1{\rm d}x_1\!\!\ldots\!\! \int\limits_0^1{\rm
d}x_{n-1}
\frac{ x_1^{\alpha_n-1}\ldots x_{n-1}^{\alpha_{n-1}-1}
(1{-}x_1{-}\ldots{-}x_{n-1})
^{\alpha_n-1}}
{[x_1a_1{{+}}\ldots{{+}}x_{n-1}a_{n-1}{{+}}(1{-}x_1{-}\ldots{-}x_{n-1})a_n]^
{\alpha_1{+}\ldots{+}\alpha_n}}\,,
\label{param}
\end{eqnarray}
where the Feynmann variables $x_i$ extend over the domain $x_1+\ldots+x_{n-1}\leqslant 1$.

%\end{widetext}
To compute the $d$-dimensional integrals we apply :
\begin{eqnarray}
&&\int
\frac{{\rm d} {p}} {(p^2{+}2\vec{k}\vec{p}{+}m^2)^{\alpha}}=
\frac{1}{2}\frac{\Gamma(d/2)\Gamma(\alpha-d/2)} {\Gamma(\alpha)}
(m^2-k^2)^{d/2-\alpha}, \label{int}
\end{eqnarray}
here ${\rm d}p={\rm d}^dp\, \Omega_d/(2\pi)^d$,  where the geometrical angular factor $\Omega_d=1/(2^{d-1}\pi^{d/2}\Gamma(d/2))$
is separated out and absorbed by redefining the coupling constant.

As an example we present the calculation of the integral:
\begin{equation}
I_2(0, q_1\,, q_2)=\int\frac{{\rm d}{\vec p}}{(p^2{+}1)[(p{+}q_1)^2{+}1][(p{+}q_2)^2{+}1]}\,.
\end{equation}
First,  we make use of formula (\ref{param}) to rewrite:
\begin{equation}
\frac{1} {(p^2{+}1)[(p{+}q_1)^2{+}1][(p{+}q_2)^2{+}1]}=
\frac{\Gamma(3) \int_0^1{\rm d}x_1  \int_0^1 {\rm d}x_2 } {[p^2{+}2\vec{p}(\vec{q}_1x_1{+}\vec{q}_2x_2) {+}1{+}x_1q_1^2{+}x_2q_2^2]^3}\,.
\nonumber
\end{equation}
Now one can perform the integration over $p$,  passing to the
$d$-dimensional polar coordinates and making use of the formula
(\ref{int}):
\begin{eqnarray}&&{
\int\frac{{\rm d}{p}}{[p^2{+}2\vec{p}(\vec{q}_1x_1{+}\vec{q}_2x_2) {+}1{+}x_1q_1^2{+}x_2q_2^2]^3}= \nonumber}\\
&& \frac{\Gamma(d/2)\Gamma(3-d/2)}
{2\Gamma(3)}[1{+}x_1q_1^2(1-x_1){+}x_2q_2^2(1-x_2)-2x_1x_2q_1q_2]^{d/2-3}.
\end{eqnarray}
As a result,  we are left with:
\begin{eqnarray}\lefteqn{
I_2(0,q_1\,,q_2)=\frac{1}{2} {\Gamma\left(\frac{d}{2}\right)}{\Gamma\left(3-\frac{d}{2}\right)}\times}\nonumber\\
&&\mbox{}\times\int\limits_0^1{\rm
d}x_1 \int\limits_0^1{\rm
d}x_2
[1{+}x_1q_1^2(1-x_1){+}x_2q_2^2(1-x_2)-2x_1x_2q_1q_2]^{d/2-3}.\label{inn}
\end{eqnarray}
To find the contributions of this integral to $G_{xx|xx}$ and $G_{xx|yy}$ (according to (\ref{Gij})),
we first differentiate the integrand in (\ref{inn})
over the components of vectors $\vec{q}_1$\,,  $\vec{q}_2$:
\begin{eqnarray*}
 I_{xx|xx}\hspace{-0.5cm}&&\equiv\frac{\rd}{\rd q_1^x}\frac{\rd}{\rd q_1^x}\frac{\rd}{\rd q_2^x}\frac{\rd}{\rd q_2^x}\left[1{+}x_1q_1^2(1-x_1){+}x_2q_2^2(1-x_2)-2x_1x_2q_1q_2\right]^{d/2-3}
\Big|_{\vec{q}_1=\vec{q}_2=0}\nonumber\\
&&=4\left(\frac{d}{2}-3\right)^{2}\left[2x_1^{2}x_2^{2}{+}x_2x_1(1-x_2)(1-x_1)\right] -4\left(\frac{d}{2}-3\right)\left[x_1x_2( 1-x_1)(1-x_2){+}4x_1^{2}x_2^{2}\right],\nonumber\\
I_{xx|yy}\hspace{-0.5cm}&&\equiv\frac{\rd}{\rd q_1^x}\frac{\rd}{\rd q_1^x}\frac{\rd}{\rd q_2^y}\frac{\rd}{\rd q_2^y}\left[1{+}x_1q_1^2(1-x_1){+}x_2q_2^2(1-x_2)-2x_1x_2q_1q_2\right]^{d/2-3}|_{\vec{q}_1=\vec{q}_2=0}
\nonumber\\
&&=\left( \frac{d}{2}-3\right )^{2} x_2x_1(1-x_2)(1-x_1)  -4\left(\frac{d}{2}-3\right)x_2x_1(1-x_1)( 1-x_2).  \nonumber
\end{eqnarray*}
Finally,  the contributions of $I_2(0, q_1\,, q_2)$ to $G_{xx|xx}$ and $G_{xx|yy}$
are found by performing integration of
$I_{xx|xx}$ and $I_{xx|yy}$ over $x_1$\,,
$x_2$ in (\ref{inn}). Results can further be evaluated either fixing the value of
space dimension $d$ or performing an expansion in parameter $\varepsilon=4-d$.
Working within the last approach,  we
obtain up to the first order of the $\varepsilon$-expansion:
\begin{eqnarray}
&&\frac{1}{2} {\Gamma\left(\frac{d}{2}\right)}{\Gamma\left(3-\frac{d}{2}\right)}
\int\limits_0^1{\rm
d}x_1 \int\limits_0^1{\rm
d}x_2\, I_{xx|xx}\simeq \frac{1}{9}{+}\frac{1}{36}\varepsilon, \nonumber\\
&&\frac{1}{2} {\Gamma\left(\frac{d}{2}\right)}{\Gamma\left(3-\frac{d}{2}\right)}
\int\limits_0^1{\rm
d}x_1 \int\limits_0^1{\rm
d}x_2\, I_{xx|yy}\simeq1{+}\frac{1}{4}\varepsilon.
\end{eqnarray}

\section*{Appendix B \label{B}}
\renewcommand{\theequation}{B\arabic{equation}}
\setcounter{equation}{0}

In this appendix we give the expressions for renormalized Green functions
$ G_{\rm R}^{(2,4)}$ with four insertions $\varphi^2({q}_1)/2$, $\varphi^2(-{q}_1)/2$, $\varphi^2({q}_2)/2$, $\varphi^2(-{q}_2)/2$
one needs for calculation of the averaged asphericity ratios $\hat{A}_d^{\rm SAW}$,
$\hat{A}_d^{\rm LR}$ defined by~(\ref{adfinal}).
The Green function $G_{\rm R}^{(2,4)}(u)$  (see figure~\ref{g24}) reads:
\begin{eqnarray}
&& G_{\rm R}^{(2,4)}(u)=\frac{8}{(q_1^2{+}1)(q_2^2{+}1)}{+}
\frac{8}{(q_1^2{+}1)(q_2^2{+}1)[(q_1{+}q_2)^2{+}1]}{+}\frac{4}{(q_1^2{+}1)^2}
{+}\frac{4}{(q_2^2{+}1)^2}\label{g24pure}\nonumber\\
&&-\frac{8}{3}\left\{ \frac{u [2I_1(0,q_1){+}2I_1(0,q_2)]}{(q_1^2{+}1)(q_2^2{+}1)[(q_1{+}q_2)^2{+}1]}-\frac{u [2I_1(0,q_1){+}2I_1(0,q_2){+}2I_2(0,q_1,q_2)]}{(q_1^2{+}1)(q_2^2{+}1)}\right.\nonumber\\
&& -\frac{2 u I_1(0,q_1)}{(q_1^2{+}1)^2[(q_1{+}q_2)^2{+}1]}-
\frac{2 u I_1(0,q_2)}{(q_2^2{+}1)^2[(q_1{+}q_2)^2{+}1]}-
\frac{1}{2}\frac{uI_2(0,0,q_2)}{(q_1^2{+}1)^2}-
\frac{1}{2}\frac{u I_2(0,0,q_1)}{(q_2^2{+}1)^2}\nonumber\\
&&-\frac{u
[I_2(0,0,q_1){+}I_3(0,0,q_1,q_2){+}I_3(0,q_1,q_2,q_1{+}q_2){+}
I_3(0,q_1,q_1,q_1{+}q_2)]}{(q_2^2{+}1)}\nonumber\\
&&-\frac{u
[I_{2}(0,0,q_2){+}I_3(0,0,q_1,q_2){+}I_3(0,q_1,q_2,q_1{+}q_2)
{+}I_3(0,q_2,q_2,q_1{+}q_2)]}{(q_1^2{+}1)}\nonumber\\
&&-\frac{u [I_2(0,q_1,q_1{+}q_2){+}I_2(0,q_2,q_1{+}q_2)]}{(q_2^2{+}1)[(q_1{+}q_2)^2{+}1]}-\frac{u [I_2(0,q_1,q_1{+}q_2){+}I_2(0,q_2,q_1{+}q_2)]}{(q_1^2{+}1)[(q_1{+}q_2)^2{+}1]}\nonumber\\
&&-\left.\frac{u
[2I_4(0,0,0,q_1,q_2){+}2I_4(0,0,q_1,q_2,q_1{+}q_2)
{+}I_4(0,0,q_1,q_1,q_1{+}q_2)
{+}I_4(0,0,q_2,q_2,q_1{+}q_2)]}{2} \right\}\nonumber\\
%\left.-\frac{6uI_8(q_1,q_2)}{2} \right\}
&&
%{+}\frac{ u(2I_{10}(q_1,q_2){+}2I_{11}(q_1,q_2{+}I_{12}(q_1|(q_1,q_2){+}I_{12}(q_2|(q_1,q_2))))}{3}
{+}\frac{4uI_1(0,0)}{3}\left[\frac{8}{(q_1^2{+}1)(q_2^2{+}1)}{+}
\frac{8}{(q_1^2{+}1)(q_2^2{+}1)[(q_1{+}q_2)^2{+}1]}
{+}\frac{4}{(q_1^2{+}1)^2}{+}\frac{4}{(q_2^2{+}1)^2}\right].
\end{eqnarray}

The Green function $G_{\rm R}^{(2,4)}(u,w)$  reads:
\begin{eqnarray*}
&& G_{\rm R}^{(2,4)} (u,w)=\frac{8}{(q_1^2{+}1)(q_2^2{+}1)}{+}
\frac{8}{(q_1^2{+}1)(q_2^2{+}1)[(q_1{+}q_2)^2{+}1]}{+}\frac{4}{(q_1^2{+}1)^2}
\label{g24LR}\nonumber\\
&&-\frac{8}{3}\left\{ \frac{u [2I_1(0,q_1){+}2I_1(0,q_2)]}{(q_1^2{+}1)(q_2^2{+}1)[(q_1{+}q_2)^2{+}1]}-\frac{u [2I_1(0,q_1){+}2I_1(0,q_2){+}2I_2(0,q_1,q_2)]}{(q_1^2{+}1)(q_2^2{+}1)}\right.\nonumber\\
&&
-\frac{w [2J_1(0,q_1){+}2J_1(0,q_2)]}{(q_1^2{+}1)(q_2^2{+}1)[(q_1{+}q_2)^2{+}1]}-\frac{w [2J_1(0,q_1){+}2J_1(0,q_2){+}2J_2(0,q_1,q_2)]}{(q_1^2{+}1)(q_2^2{+}1)}\nonumber\\
&& -\frac{2 uI_1(0,q_1)}{(q_1^2{+}1)^2[(q_1{+}q_2)^2{+}1]}-
\frac{2 u I_1(0,q_2)}{(q_2^2{+}1)^2[(q_1{+}q_2)^2{+}1]}-
\frac{1}{2}\frac{uI_2(0,0,q_2)}{(q_1^2{+}1)^2}-
\frac{1}{2}\frac{u I_2(0,0,q_1)}{(q_2^2{+}1)^2}\nonumber\\
&& -\frac{2 w J_1(0,q_1)}{(q_1^2{+}1)^2[(q_1{+}q_2)^2{+}1]}-
\frac{2 w J_1(0,q_2)}{(q_2^2{+}1)^2[(q_1{+}q_2)^2{+}1]}-
\frac{1}{2}\frac{w J_2(0,0,q_2)}{(q_1^2{+}1)^2}-
\frac{1}{2}\frac{w J_2(0,0,q_1)}{(q_2^2{+}1)^2}\nonumber\\
&&-\frac{u
[I_2(0,0,q_1){+}I_3(0,0,q_1,q_2){+}I_3(0,q_1,q_2,q_1{+}q_2){+}
I_3(0,q_1,q_1,q_1{+}q_2)]}{(q_2^2{+}1)}\nonumber
\end{eqnarray*}
\begin{eqnarray}
&&-\frac{w
[J_2(0,0,q_1){+}J_3(0,0,q_1,q_2){+}J_3(0,q_1,q_2,q_1{+}q_2){+}
J_3(0,q_1,q_1,q_1{+}q_2)]}{(q_2^2{+}1)}\nonumber
\\
&&-\frac{u
[I_{2}(0,0,q_2){+}I_3(0,0,q_1,q_2){+}I_3(0,q_1,q_2,q_1{+}q_2)
{+}I_3(0,q_2,q_2,q_1{+}q_2)]}{(q_1^2{+}1)}\nonumber\\
&&-\frac{w
[J_{2}(0,0,q_2){+}J_3(0,0,q_1,q_2){+}J_3(0,q_1,q_2,q_1{+}q_2)
{+}J_3(0,q_2,q_2,q_1{+}q_2)]}{(q_1^2{+}1)}\nonumber
\\
&&-\frac{u [I_2(0,q_1,q_1{+}q_2){+}I_2(0,q_2,q_1{+}q_2)]}{(q_2^2{+}1)[(q_1{+}q_2)^2{+}1]}-\frac{u [I_2(0,q_1,q_1{+}q_2){+}I_2(0,q_2,q_1{+}q_2)]}{(q_1^2{+}1)[(q_1{+}q_2)^2{+}1]}\nonumber
\\
&&-\frac{w [J_2(0,q_1,q_1{+}q_2){+}J_2(0,q_2,q_1{+}q_2)]}{(q_2^2{+}1)[(q_1{+}q_2)^2{+}1]}-\frac{w [J_2(0,q_1,q_1{+}q_2){+}J_2(0,q_2,q_1{+}q_2)]}{(q_1^2{+}1)[(q_1{+}q_2)^2{+}1]}\nonumber
\\
&&-\frac{u [2I_4(0,0,0,q_1,q_2){+}2I_4(0,0,q_1,q_2,q_1{+}q_2)
{+}I_4(0,0,q_1,q_1,q_1{+}q_2)
{+}I_4(0,0,q_2,q_2,q_1{+}q_2)]}{2}\nonumber
\\
&&-\left.\frac{w
[2J_4(0,0,0,q_1,q_2){+}2J_4(0,0,q_1,q_2,q_1{+}q_2)
{+}J_4(0,0,q_1,q_1,q_1{+}q_2)
{+}J_4(0,0,q_2,q_2,q_1{+}q_2)]}{2} \right\}\nonumber\\
&&
%{+}\frac{ u(2I_{10}(q_1,q_2){+}2I_{11}(q_1,q_2{+}I_{12}(q_1|(q_1,q_2){+}I_{12}(q_2|(q_1,q_2))))}{3}
{+}\frac{4[uI_1(0,0) - w
J_1(0,0)]}{3}\left[\frac{8}{(q_1^2{+}1)(q_2^2{+}1)}{+}
\frac{8}{(q_1^2{+}1)(q_2^2{+}1)[(q_1{+}q_2)^2{+
}1]}{+}\frac{4}{(q_1^2{+}1)^2}{+}\frac{4}{(q_2^2{+}1)^2}\right].\nonumber\\
\end{eqnarray}

The one-loop integrals  $I_i$ and $J_i$ are explained in appendix~A.

\ukrainianpart

\title{Форми макромолекул у хороших розчинниках: \\ підхід теоретико-польової ренормалізаційної групи}
\author{В. Блавацька\refaddr{label1}, К. фон Фербер\refaddr{label2,label3}, Ю. Головач\refaddr{label1}}
\addresses{
\addr{label1} Інститут фізики конденсованих систем НАН України,  вул. І. Свєнціцького, 1, 79011 Львів, Україна
\addr{label2} Дослідницький центр прикладної математики, Університет Ковентрі, CV1 5FB Ковентрі, Англія
\addr{label3} Теоретична фізика полімерів, Університет Фрайбургу, D-79104 Фрайбург, Німеччина
}

\makeukrtitle
\begin{abstract}
\tolerance=3000%

У статті ми показуємо, яким чином можна застосувати метод теоретико-польової ренормалізаційної групи
для аналізу універсальних властивостей форм довгих гнучких полімерних ланцюгів у пористому середовищі.
До цього часу такі аналітичні розрахунки в основному торкались показників скейлінгу, що визначають  конформаційні
властивості полімерних макромолекул. Проте, існують й інші спостережувані величини, що, як і показники скейлінгу,
 є універсальними  (тобто незалежними від хімічної структури як макромолекул, так і розчинника), а отже можуть бути проаналізовані в межах підходу ренормалізаційної групи.  Ми цікавимось питанням, якої форми набуває довга гнучка полімерна макромолекула у розчині в присутності пористого середовища. Це питання є суттєвим для розуміння поведінки  макромолекул у колоїдних розчинах, поблизу мікропористих мембран, а також у клітинному середовищі.
Ми розглядаємо запропоновану раніше модель полімера у $d${-}вимірному просторі
[V.~Blavats'ka, C.~von Ferber, Yu.~Holovatch, Phys. Rev.~E, 2001,  {\bf 64}, 041102] у середовищі
із структурними неоднорідностями, що характеризуються парною кореляційною  функцією $h(r)$,
яка спадає із відстанню $r$ згідно степеневого закону:  $h(r) \sim r^{{-}a}$. Застосовуємо
підхід теоретико-польової ренормалізаційної групи і оцінюємо відношення розмірів
$\langle R_{\rm e}^2 \rangle/\langle R_{\rm G}^2 \rangle$ та асферичність $\hat A_d$
до першого порядку  $\varepsilon=4{-}d$, $\delta=4{-}a$-розкладу.
\keywords полімер, заморожений безлад, ренормалізаційна група

\end{abstract}
\end{document}